\documentclass[12pt]{iopart}

\pdfoutput=1

\usepackage{graphicx}
\usepackage{iopams}

\newcommand{\omd}{\omega_\mathrm{d}}
\newcommand{\oma}{\omega_\mathrm{a}}
\newcommand{\cR}{c_\mathrm{R}}
\newcommand{\kd}{k_\mathrm{d}}
\newcommand{\kp}{k_\mathrm{p}}

\newcommand{\rhoup}{\rho_\mathrm{u,+}}
\newcommand{\rhoum}{\rho_\mathrm{u,-}}
\newcommand{\rhobp}{\rho_\mathrm{b,+}}
\newcommand{\rhobm}{\rho_\mathrm{b,-}}
\newcommand{\rhobpm}{\rho_\mathrm{b,\pm}}
\newcommand{\rhoupm}{\rho_\mathrm{u,\pm}}
\newcommand{\rhou}{\rho_\mathrm{u}}
\newcommand{\qc}{q_\mathrm{c}}
\newcommand{\Jb}{J_\mathrm{b}}

\begin{document}

\title[Particle interactions and lattice dynamics for efficient transport]{Particle interactions and lattice dynamics: Scenarios for efficient bidirectional stochastic transport?}

\author{M Ebbinghaus$^{1,2,3}$, C Appert-Rolland$^{1,2}$ and L Santen$^3$}

\address{$^1$ Univ. Paris-Sud, Laboratoire de Physique Th\'eorique, B\^at. 210, F-91405 Orsay Cedex, France}
\address{$^2$ CNRS, LPT, UMR 8627, B\^at 210, F-91405 Orsay Cedex, France}
\address{$^3$ Fachrichtung Theoretische Physik, Universit\"at des Saarlandes, D-66123 Saarbr\"ucken, Germany}
\ead{\mailto{ebbinghaus@lusi.uni-sb.de}, \mailto{cecile.appert-rolland@th.u-psud.fr}, \mailto{l.santen@mx.uni-saarland.de}}

\begin{abstract}
Intracellular transport processes driven by molecular motors
can be described by stochastic lattice models of self-driven
particles. Here we focus on bidirectional transport models
excluding the exchange of particles on the same track.  We
explore the possibility to have efficient transport in these
systems.  One possibility would be to have appropriate
interactions between the various motors' species, so as to form
lanes.  However, we show that the lane formation mechanism
based on modified attachment/detachment rates as it was
proposed previously is not necessarily connected to an
efficient transport state and is suppressed when the
diffusivity of unbound particles is finite.  We propose another
interaction mechanism based on obstacle avoidance that allows
to have lane formation for limited diffusion.  Besides, we had
shown in a separate paper that the dynamics of the lattice
itself could be a key ingredient for the efficiency of
bidirectional transport.
Here we show that lattice dynamics and interactions can both
contribute in a cooperative way to the efficiency of transport.
In particular, lattice dynamics can decrease the interaction
threshold beyond which lanes form.  Lattice dynamics may also
enhance the transport capacity of the system even when lane
formation is suppressed.
\end{abstract}


\maketitle

\section{Introduction}
Many biological processes depend on efficient transport on the microscopic scale. Biological cells contain a large number of microscopic machines that are for example able to maintain the exchange of nutrients between the cell and the surrounding medium. Another important issue is the transport of cargos between the cell center and the membrane~\cite{Alberts02}. In this context, transport driven by molecular motors is of particular importance: Molecular motors are specialized proteins that are able to perform a directed motion along actin filaments and microtubules, i.e., filaments that are part of the cytoskeleton~\cite{Schliwa03}. 

Several aspects of motor driven biological transport have been modeled recently~\cite{Lipowsky01,Parmeggiani03,Klumpp03,Evans03,Juhasz04,Parmeggiani04,Tailleur09} with variants of the so-called asymmetric exclusion process (ASEP) \cite{Krug91,Derrida98a}. The use of this type of models has been motivated since the molecular motors move stepwise (hence the relevance of a model defined on a discrete lattice) and interact via mutual exclusion. But in addition to this, molecular motors can attach and detach from the filament. This property of molecular motors led to extensions of the ASEP~\cite{Lipowsky01,Parmeggiani03,Parmeggiani04} which successfully describe the unidirectional motion of \emph{in vitro} motor assays \cite{Nishinari05}.

In the case of bidirectional systems of molecular motors, the theoretical understanding is so far incomplete. It has been shown that simple extensions of the unidirectional models are not able to describe efficient bidirectional transport if the filament is embedded in a small volume and the diffusivity of unbound particles is limited~\cite{Ebbinghaus09}. This situation could for example be relevant for motor-driven transport in axons. The question that naturally arises is the following:
What are the minimal ingredients that must be added to the existing unidirectional models in order to describe bidirectional transport of stochastic interacting particles?  

In this article, we consider different kinds of
particle-particle interactions which may lead to lane
formation, i.e., the self-organized separation of particles
moving in opposite directions. This is in contrast to models for bidirectional transport with predefined lanes for each direction (e.g.,~\cite{juhasz07b}). 
We shall discuss their 
ability to generate efficient bidirectional
transport.
A first mechanism which has
formerly been proposed by Klumpp and
Lipowsky~\cite{Klumpp04} takes into account the
experimentally observed preferential adsorption of motors
to a filament if the neighboring site is occupied by a
motor of the same type~\cite{Vilfan01,Roos08,Muto05}.
In~\cite{Klumpp04}, they have observed lane formation above
a critical value of the coupling parameter. We complete
their results and present a more extensive study of the
model.
We show that the properties of the model depend strongly
on the kind of diffusion reservoir that is considered
around the filaments.
In particular we have shown that lane formation is hindered if
diffusion occurs in a confined environment.
In a second part, as an alternative mechanism, we introduce particle-particle interactions that induce lane changes if oppositely moving particles are encountered.
This mechanism does allow to have lane formation in a confined
environment.

Eventually, we also consider that the filament itself
can undergo some dynamical process, motivated by the
experimentally observed continuous reorganization of the
microtubule network. We shall consider a much simplified
version of the filament dynamics, as this was done
in~\cite{Ebbinghaus10}. There, we had shown that a
dynamic lattice may enhance the transport efficiency of
bidirectional stochastic transport. Here, we shall
study the influence of lattice dynamics in the presence of
particle-particle interactions and show that both
mechanisms can cooperate to obtain efficient transport.

Our work is organized as follows: After introducing some general
definitions and summarizing some previously obtained results
(next section), we
shall consider two types of interactions: first, we
investigate the model introduced by Klumpp and Lipowsky~\cite{Klumpp04}
considering the preferential binding of particles
(section~\ref{sec:interact}). In section~\ref{sec:sidestep},
we discuss the effect of another type of interaction,
namely a dynamic particle-particle interaction on the
filament inducing lane changes. As mentioned above, for both types of
interactions, we apply different kinds of particle reservoirs: we shall consider grand-canonical and canonical
particle reservoirs and also the effect of a finite
diffusivity of unbound particles.
We shall also systematically
discuss the impact of a dynamic lattice on lane formation and transport
capacity of the systems. In the final section, we compare the different
approaches and discuss their relevance for axonal  transport.

\section{General definitions}

\subsection{Model definition}
\label{subsec:modeldef}

The models which are discussed in this article are variants of the totally asymmetric
simple exclusion process (TASEP) with two species of particles. The two species
have opposite `charges' resulting in opposite moving directions on a
one-dimensional lattice with regular spacing. The one-dimensional lattice
represents the microtubule, or more precisely, one of the filaments that
compose the microtubule, and thus it will be referred to as a filament in the
remaining of the paper.

\begin{figure}[tbp]
  \begin{center}
    \includegraphics[scale=0.8, clip]{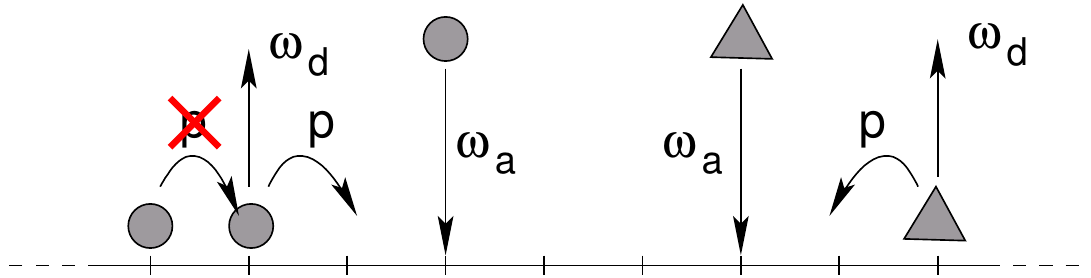}
    \caption{Sketch of the hopping rules of the two different particle species, depicted by circles and triangles. Particles step forward at rate $p$ if the target site is empty. A particle detaches from the lattice at rate $\omd$ and adsorbs to unoccupied sites at rate $\oma$.}
    \protect\label{fig:basic_model}
  \end{center}
\end{figure}

 Particles hop with rate $p$ along the filament in the direction determined by
 their charge which we will refer to as positive or negative depending on the
 preferential moving direction
(see \fref{fig:basic_model}).
 We additionally consider attachment and detachment of particles to and from the lattice as first proposed in~\cite{Lipowsky01}.
 Particles detach from the lattice with rate $\omd$ to a reservoir. The adsorption of a
particle from the reservoir onto a site of the lattice occurs at rate $\oma$.
We examine different kinds of reservoirs which are specified below in the
corresponding sections. They have in common that particles in the reservoir
diffuse freely without any interactions between particles. 
On the filament, all particles
interact via hard-core exclusion and positional exchanges of oppositely charged particles are excluded in contrast to the model considered in~\cite{Arndt98}.  A stepping or attachment move
is therefore rejected if the target site is already occupied by another
particle.

As standard set of parameters for the
particle dynamics, we will use $p=1$, $\oma=0.1$ and $\omd=0.01$. This choice
of parameters corresponds to processive motion (i.e., motors are attached on the
filament long enough so that directed motion contributes significantly to
the overall motion), a
basic feature of the molecular motors contributing to long-range transport, as conventional kinesin or cytoplasmic dynein~\cite{Alberts02}.

As there are several parallel microtubules in axons, each of them
made of several protofilaments, we could a priori consider a large number $n_\mathrm{fil}$
of parallel filaments. However, in order to keep the model simple,
we shall consider only one filament except when the mechanism under
study requires to have at least two filaments in parallel.

Particle densities are denoted by $\rho_\pm$ and defined as the number of particles divided by the system length. The index refers to the particle species. Densities and particle numbers are therefore related through
$N_\pm=\rho_\pm\times L\times n_\mathrm{fil}$ and we choose the total
density to be equal to $\rho = 2\rho_+ = 2\rho_-$ throughout the paper.
Note that $\rho$ and $\rho_\pm$ can be greater than one as they take into account the total number of
particles (including those in the reservoir which is assumed to have an infinite capacity in this paper), divided by the \emph{number of
filament sites}. The quantities $\rho_\mathrm{b}$, $\rho_{\mathrm{b},\pm}$ denote the densities of bound particles on the filament and $\rho_\mathrm{u}$, $\rho_{\mathrm{u},\pm}$ the densities of particles in the reservoir.

\subsection{Reservoirs}
\label{subsec:reservoir_def}

When motors detach from the microtubule, they undergo diffusive
motion in the surrounding cytoplasm - a region of the cell that
will be referred to as the \emph{reservoir}.
Depending on the geometry of the reservoir, the diffusive motion
can have various characteristics. We shall define three types of
reservoirs below.

\subsubsection{Grand-canonical reservoir}

If the surroundings of the filament is an open space, motors
can diffuse very far from the filament, and thus, most of the
time, motors that attach to the filament have completely lost
the memory of their previous detachment point.
Hence there is no need to keep track of spatial positions of
the diffusing motors.

Actually, the grand-canonical reservoir will be modeled by a single
number, representing the density of motors in the reservoir, which
is assumed to be constant and (in this paper) equal to
$\rhou=2\rhoup=2\rhoum=0.1$.
Attachment to empty sites on the filament occurs at rate $\rhoupm\cdot\oma$.

This type of reservoir is relevant if the filament is embedded in a large volume and the number of attached motors is small compared to the number of motors in solution. This is the case for many \emph{in vitro} experiments.

\subsubsection{Canonical reservoir}

An infinite diffusion rate in the surroundings of the lattice
is also considered in the canonical reservoir, and thus
particles in the reservoir do not
have any spatial coordinate and can attach to any empty site along the lattice.

However, in the canonical setup, we consider that the total number of
particles in the system is finite. This implies, with our definition of
the density, that 
$\rho_\pm=\rhobpm+\rhoupm$
is fixed.
Again, attachment occurs at rate $\rhoupm\cdot\oma$.

This type of reservoir is relevant if the filament is embedded in a large volume and the number of motors attached to the filament is of the same order as the number of motors in solution. This situation can also be encountered in \emph{in vitro} experiments.

\subsubsection{Finite diffusion}
\label{subsubsec:finite_diff}

Actually, the axon is a rather crowded environment\footnote{
The importance of confinement in the axon was for example
illustrated by the simulations in \cite{greulich_s11}
which reproduce the experimental results of H. Erez et al
\cite{erez07} for transected axons.
},
where the size of the transported objects is of the
same order as the distance between the microtubules
\cite{shemesh_s10b,chen92}, and the previous reservoirs are not relevant to model
the diffusive motion episodes.
In order to study the effects of confinement, we shall introduce
a reservoir with a memory effect: 
there must be a correlation between the locations where a motor detaches,
and where it attaches again on the microtubule.

\begin{figure}[tbp]
  \begin{center}
    \includegraphics[scale=0.8, clip]{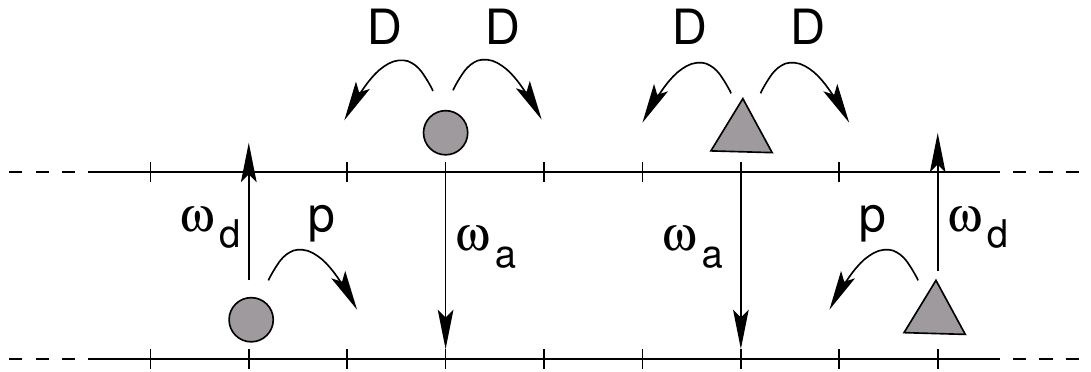}
    \caption{Schematic representation of the particle dynamics in the system with finite diffusion in the reservoir. The lower lane represents the filament already shown in \fref{fig:basic_model}. The upper lane is the reservoir within which particle do not interact with each other but diffuse at finite rate $D$.
    }
    \protect\label{fig:finitediff}
  \end{center}
\end{figure}

In order to do so, the model is extended by a
second lane which represents the reservoir
(see \fref{fig:finitediff}). Particles in the reservoir
symmetrically hop to both neighboring sites with rate $D$ without any
interaction between particles. Note that this
one-dimensional reservoir actually represents a three-dimensional region
around the microtubule, which allows for several particles at the
same longitudinal position.
Only the longitudinal diffusion, along the microtubule, is modelled
explicitly.
The transversal dimensions of the three-dimensional diffusion reservoir,
which are not represented explicitly,
can be tuned by taking a larger or weaker effective attachment rate (see
\cite{Ebbinghaus10b} for an example). The larger the diffusive region is
(in the direction transverse to the microtubule),
the weaker the effective attachment rate is.
Thus, under strong confinement, diffusion in the transverse direction
should be more limited, the attachment rate in the model will be larger,
and the memory effect will increase.

\subsection{Lattice dynamics}

\subsubsection{Definition}
\label{subsec:latticedyn_def}

\begin{figure}[tbp]
  \begin{center}
    \includegraphics[scale=0.8, clip]{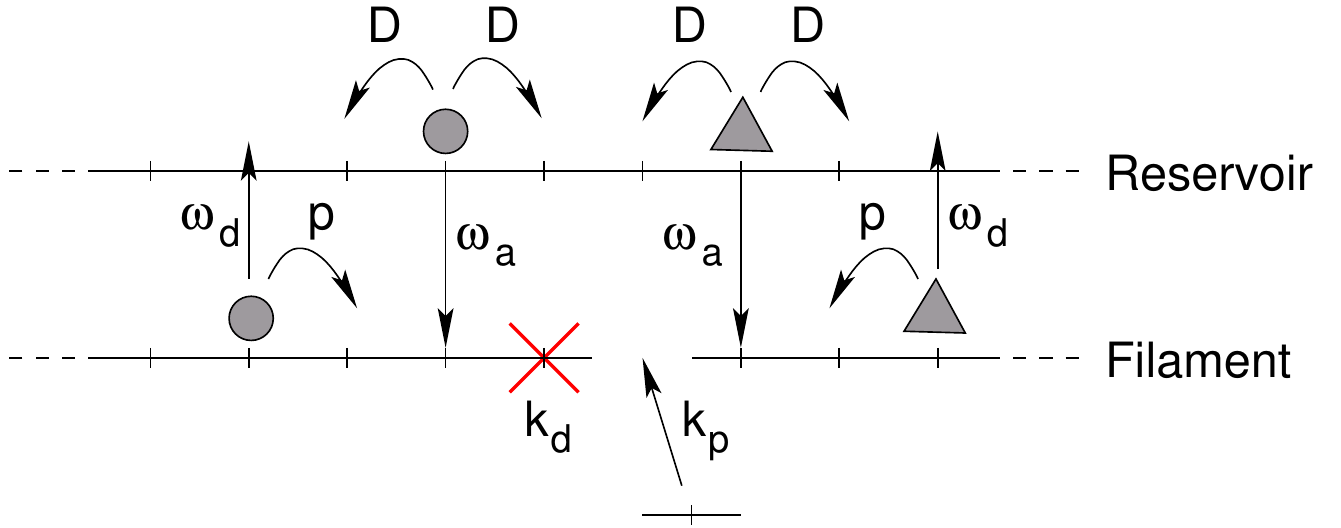}
    \caption{Same figure as in Fig. \protect{\ref{fig:finitediff}},
    but the lattice dynamics has been added.
    A site of the filament (lower lane) can be suppressed with
    rate $k_d$, and restored afterwards with a rate $k_p$.
    When a particle arrives on a site that has been suppressed,
    it goes in the diffusion reservoir (upper lane).
    }
    \protect\label{fig:model_defdyn}
  \end{center}
\end{figure}

In this paper, we shall also consider that the underlying
lattice itself can evolve dynamically.
We shall consider the same type of lattice dynamics as
 in~\cite{Ebbinghaus10}: a site of the lattice is randomly eliminated at rate
$\kd$ and recreated at rate $\kp$ (see figure \ref{fig:model_defdyn}). The indices refer to the processes of
microtubule de-/polymerization, although the dynamics we consider here have
little in common with realistic dynamics and must rather be considered as a
minimal model. A particle cannot bind to an
eliminated site and immediately detaches from the lattice if it steps onto an
eliminated site or is bound to a site which is being eliminated. We fix the
polymerization rate equal to $\kp=1$ in order to reduce the number of free
parameters (we saw indeed in~\cite{Ebbinghaus10} that transport features
were more sensitive to $\kd$ than to $\kp$).

\subsubsection{Summary of previous results - impact of the lattice dynamics}

It has been shown that bidirectional transport models in which the particle species interact only via an exclusion interaction generically exhibit macroscopic clustering at any finite density $\rho$~\cite{Ebbinghaus09}. This large cluster controls the current in the system and therefore limits the total transport capacity. In particular, for large systems, the current vanishes.

Considering the case of bidirectional transport on dynamical lattices, it has been shown that lattice dynamics can induce a transition from the clustered phase to a homogenous phase in which the particles are distributed over the whole system~\cite{Ebbinghaus10}. The homogenous states are density-dependent so that a finite (non-vanishing) current of particles is present even in very large systems, which would not be the case on a static lattice. The transition takes place because the lattice dynamics breaks up forming clusters and thus limits the reachable maximum cluster sizes. At a given strength of the lattice dynamics, the nascent clusters are dissolved more quickly than they can form and the system appears homogenous. The transition depends mostly on the rate at which binding sites disappear suggesting that the finite lifetime of individual binding sites is crucial for the observed transition.

In order to support this point of view, it has been verified that this effect is not simply due to the cutting of the filament into several segments of different lengths, by considering a static lattice with the same density of individual holes. In that case, the phase transition is not observed, thus indicating that the cluster dissolution is really an effect of
the dynamics of the lattice.

Though this scenario was inspired by the fact that 
microtubules are themselves evolving through the so-called
dynamic instability, i.e., events of polymerization and depolymerization,
the dynamics of the filament as we have defined it in~\cite{Ebbinghaus10}
or in section \ref{subsec:latticedyn_def}
do not aim at being realistic, but rather are the simplest kind
of lattice dynamics that could be considered in order to illustrate
the aforementioned mechanism.

\section{Interactions through modified attachment/detachment rates}
\label{sec:interact}

\subsection{Model definition}\label{subsec:interact_def}

\begin{figure}[tbp]
  \begin{center}
    \includegraphics[scale=0.8, clip]{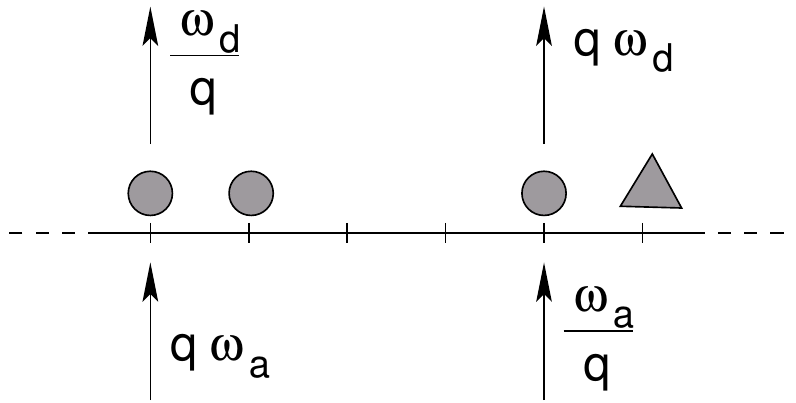}
    \caption{Modified attachment/detachment rates. The attachment (detachment) rate increases (decreases) by a factor $q$ if the next site is occupied by a particle of the same species. In the case of a particle of the opposite species, detachment is enhanced and attachment is de-favored.}
    \protect\label{fig:KL_interaction}
  \end{center}
\end{figure}

Decoration experiments with kinesin on microtubules~\cite{Vilfan01,Roos08,Muto05} showed that kinesins attach on the microtubule preferentially in the vicinity of similar motors. This motivates the assumption of a general attractive interaction between molecular motors of the same species. This experimental fact was first considered by Klumpp and Lipowsky~\cite{Klumpp04} in a modification of the above-presented basic model by consideration of detachment and attachment rates that depend on the next neighbor in the stepping direction of the particle: If the next
site in moving direction is occupied by a particle of the same species, the
detachment rate is lowered by a factor $1/q$ whereas the attachment to that
site is enhanced by a factor $q$. If the next site is occupied by a particle of
the opposite species, the detachment is enhanced by a factor $q$ and attachment
is lowered by a factor $1/q$ (see \fref{fig:KL_interaction}). In all other
cases, the rates remain unchanged.
In the remaining, we shall call $q$ the strength of the interaction,
as the modified attachment and detachment rates are supposed to be
due to attractive/repulsive interactions between similar/different
motor species.

We shall now study the consequences of having these modified
attachment/detachment rates on the efficiency of transport,
as an extension of the results presented in~\cite{Klumpp04}.
In particular, several types of reservoirs will be
considered and shown to have a strong influence on transport
properties.

In a first stage, we shall study the influence of these modified
attachment/detachment rates on a static lattice, i.e., the lattice
dynamics defined in section \ref{subsec:latticedyn_def} will not
be considered.

\subsection{Static lattice}

\subsubsection{Infinite diffusion rate}\label{subsubsec:static_infinite}
In~\cite{Klumpp04}, it has been shown that for a grand-canonical setup, lane formation is observed if the particles' attachment/detachment rates are
modified according to figure~\ref{fig:KL_interaction} on a static 
lattice.
In this section, we briefly present some new results obtained
in the same frame, in order to
discuss the efficiency of the lane formation mechanism.

\begin{figure}[tbp]
  \begin{center}
    \includegraphics[scale=0.5, clip]{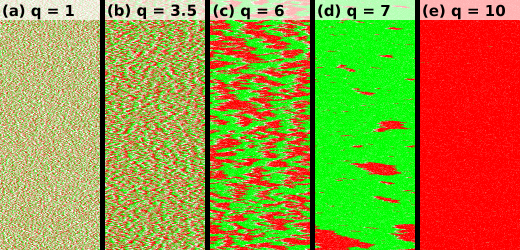}
    \caption{Space-time plots of the filament with a grand-canonical reservoir on a static filament. Every line is a snapshot of the filament of length $L=400$ for different interaction strengths $q$: (a) $q=1$, (b) $q=3.5$, (c) $q=6$, (d) $q=7$, (e) $q=10$. Red dots represent particles moving to the right, green dots are particles moving to the left, white dots are vacancies on the filament.}
    \protect\label{fig:gc_static:config}
  \end{center}
\end{figure}

\paragraph{Grand-canonical reservoir}
As already shown in~\cite{Klumpp04}, lane formation occurs in the grand-canonical setup if the interaction strength $q$ exceeds a critical value $\qc$. This effect is illustrated in the space-time plots in figure~\ref{fig:gc_static:config}. With increasing interaction strength, the typical size of the jams in the system increases, until one particle species completely takes over the system at $\qc$ so that all particles move in the same direction.

\begin{figure}[tbp]
  \begin{center}
   \includegraphics[scale=0.25, clip]{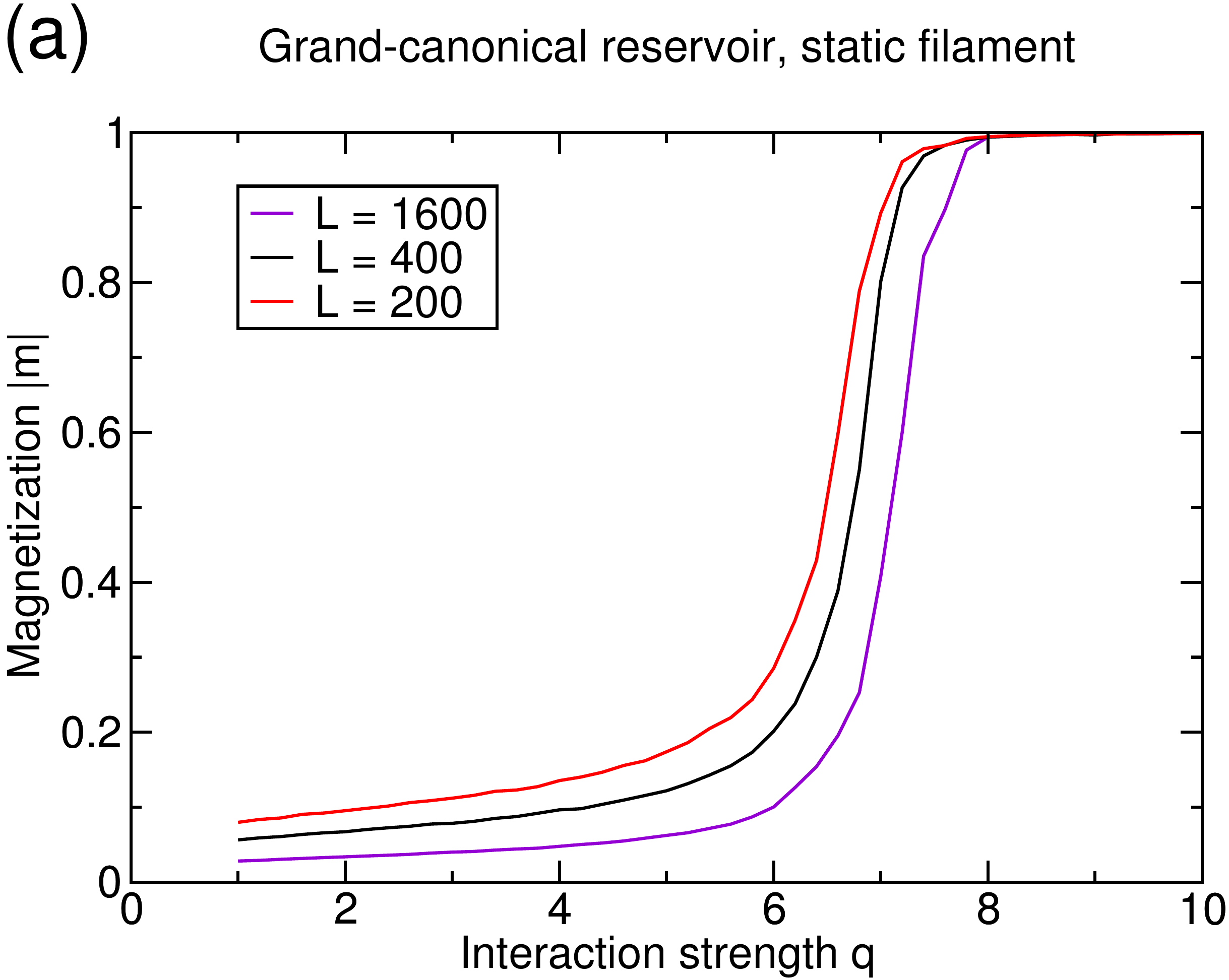}
    \includegraphics[scale=0.25, clip]{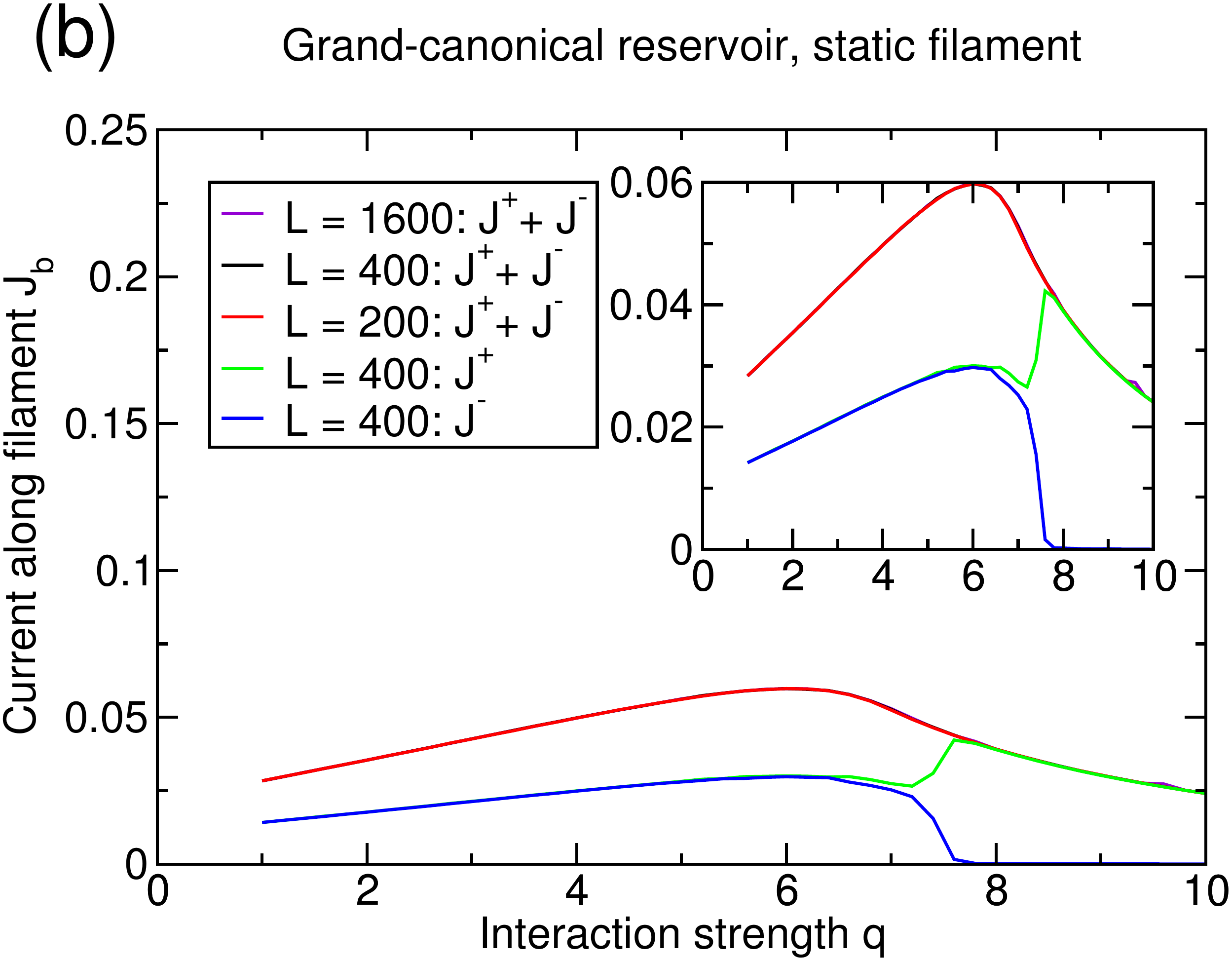}
    \caption{(a) Magnetization $m$ of the filament in systems of size
    $L=1600$ (purple), $L=400$ (black) and $L=200$ (red). (b) Current along a static
    filament of the dominant (green) and excluded (blue) particle
    species.
    The sum of both currents is shown in black and red for systems
    of size $L=1600$, $L=400$ and $L=200$ respectively, with an almost perfect
    overlap for all system sizes.
    All quantities are plotted against the interaction strength
    $q$ tuning the modified attachment/detachment rates,
    and were obtained for a grand-canonical reservoir.
    All graphs displaying the current in this article will
    have a vertical axis going to $\Jb=0.25$ which corresponds to the
    maximum current of the prototypal TASEP. The inset then always shows
    the same data on a more appropriate scale of the vertical axis if
    necessary.}
    \protect\label{fig:gc_static:current}
  \end{center}
\end{figure}

The transition manifests itself by a dramatic increase of the magnetization as shown in~\cite{Klumpp04}. Here, we define the magnetization slightly differently, since we normalize the difference in the densities of the two particle species with the total density of particles on the filament $m=(\rhobp-\rhobm)/(\rhobp+\rhobm)$. This quantity takes the value one for completely demixed subsystems, i.e., in the case of lane formation and is zero for equal numbers of particles of both species on the filament (see \fref{fig:gc_static:current}(a)). For $m\approx 0$ or $m\approx 1$, we will refer to unpolarized or polarized states, respectively, as is known from systems with magnetization. The currents of the two species along the filament exhibit a bifurcation at the transition: while one species dominates the filament and moves in a TASEP-like fashion, the excluded species does not participate in transport at all for $q>\qc$ (see \fref{fig:gc_static:current}(b)). In a real system, there are several filaments, so that, above the transition, some of them conduct transport in one direction while others conduct transport in the other direction.

Two remarks can be made from the data shown in \fref{fig:gc_static:current}(b).
First, the current does not exhibit strong finite-size
effects, although the magnetization does. In particular, the
critical interaction strength $\qc$ above which lane formation
occurs depends on the system size and increases for larger
systems. In~\cite{Klumpp04}, numerical evidence was given that
$\qc$ has a finite value in the limit of infinite system sizes
which is supported by the mean field treatment of the system in
the same paper. However, one has to be cautious before claiming
that there is true symmetry breaking in the thermodynamic limit.
Indeed, a subtlety has been exemplified in the case of
another bidirectional transport
model, the AHR model~\cite{Arndt98}, for which numerical
evidence for a phase transition had been found.
Actually, a very careful analysis~\cite{Rajewsky00} of the AHR
model has shown that the correlation length can grow up
to $10^{70}$, and then remain finite, so that there is no true
symmetry breaking.
So, although the data presented
in~\cite{Klumpp04} seem to be rather convincing, there might
be a subtlety which cannot be addressed by a numerical
treatment.

Second, although the
difference in current between the two species is very large above the
transition, indicating that lane formation has occurred, the transition does not influence the total current of the system as it is continuous. For the choice of parameters considered here, the maximum of the total current is actually located before the onset of
lane separation. This can be seen by the sum of the currents (black and red
line in \fref{fig:gc_static:current}(b)) which corresponds to the current of a
single species in a system with two filaments, of which each is dominated by a
different species. It is to note that the relative position of the onset of lane formation with respect to the maximum of the total current varies with the reservoir density such that the onset of lane formation might also occur for lower interaction strengths than the maximum in the current. The interesting feature is that, above the transition for high values of $q$, the
current actually decreases although the encounters of oppositely moving particles
are less frequent. This decrease is instead caused by the filament being overcrowded,
as can be seen in
\fref{fig:gc_static:config}. The density on the filament is close to one and
although all particles move in the same direction, most of the attempted steps
are rejected because of the lack of vacancies. The stronger the interactions
between particles, the more the situation worsens. Consequently, the current
continues to decrease for increasing interactions strengths, since the
particles have an ever-increasing affinity to the filament and the number of
vacancies thus decreases with increasing $q$.

\paragraph{Canonical reservoir}

Considering that actually, the number of motors in the system may
not be infinite,
we fix the total number of positive (negative) particles
$N_+$ ($N_-$) and observe the consequences of using a canonical
reservoir instead of a grand-canonical one on the transport properties.

Besides, we consider here more than only one track on which particles
can walk. In this scenario, choosing
more than one filament becomes important since lane formation would be hindered
otherwise. In a system with a single filament, the species which is excluded
from the filament becomes dominant in the reservoir and therefore tends to
suppress lane formation. Putting two (or more generally
any pair number of) filaments in a system resolves this issue. This is an appropriate approach as in the biological situation, one typically has several filaments in parallel. In the following, we shall consider the case of two filaments only.

Note that for multilane unidirectional systems, strong correlations between the lanes were found, both for strong and weak coupling~\cite{reichenbach_f_f06,schiffmann_a_s10}. Thus, simulating only one lane turns out to be representative for the whole system. This is in contrast to the bidirectional case considered here, for which lane formation may require to consider more than one filament.

\begin{figure}[tbp]
  \begin{center}
    \includegraphics[scale=0.25, clip]{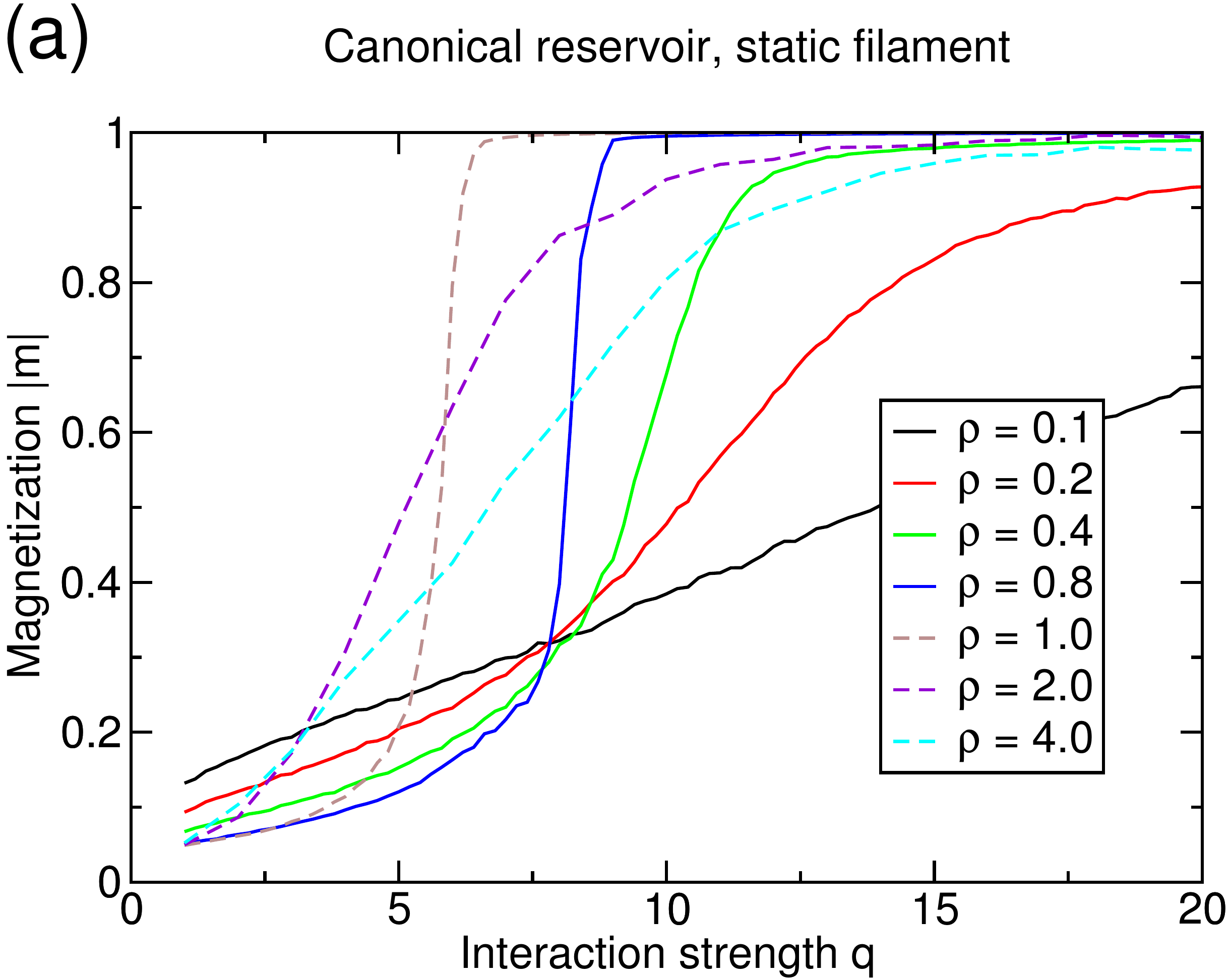}
    \includegraphics[scale=0.25, clip]{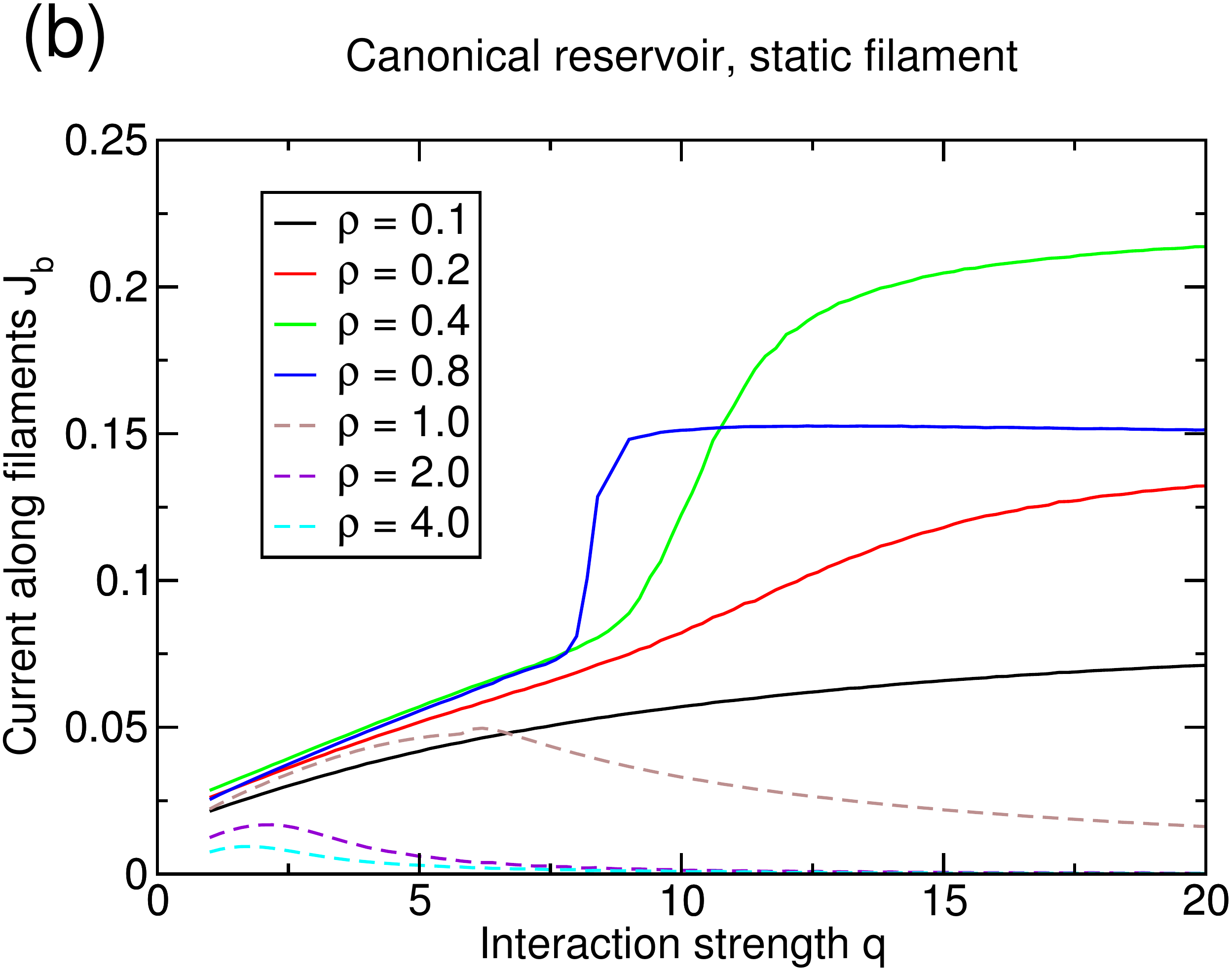}
    \caption{(a) Magnetization of the filament
    $m=(\rhobp-\rhobm)/(\rhobp+\rhobm)$ and (b) current of positive 
    particles along both
    filaments in the canonical setup with a static
    filament, for a system size $L=200$ and different densities $\rho$.
    Full lines (\full) are for densities $\rho<1$ and dashed lines
    (\dashed) are for densities $\rho\geq1$.
All quantities are plotted against the interaction strength
$q$ tuning the modified attachment/detachment rates.}
    \protect\label{fig:c_static:magnetization_current}
  \end{center}
\end{figure}

The behavior of the system is qualitatively different for densities
$\rho=2\rho_+=2\rho_-\geq 1$ and $\rho<1$ since in the latter case, there are
not enough particles to fill both filaments completely with a single
particle species each. The results are illustrated in
\fref{fig:c_static:magnetization_current}. For the magnetization, it is
interesting to see that the transition still takes place, even at very low
total particle densities. The fewer particles are present in the system, the
higher the interaction between particles has to be in order to induce
polarization. For $\rho = 1$, there are exactly as many particles of one
species in the system as there are sites on a single filament and lane
formation is extremely stable in this case since there are enough particles to
mediate the interaction at best but not too many to avoid that particles frequently
attach to the ``wrong'' filament on which the species is not dominant.
For densities $\rho>1$, the latter effect is not avoided anymore as the reservoir cannot be emptied.
There are always many particles of both species in the reservoir which
interfere with the existing lanes such that the transition to the polarized
state is hindered.

\begin{figure}[tbp]
  \begin{center}
    \includegraphics[scale=0.25, clip]{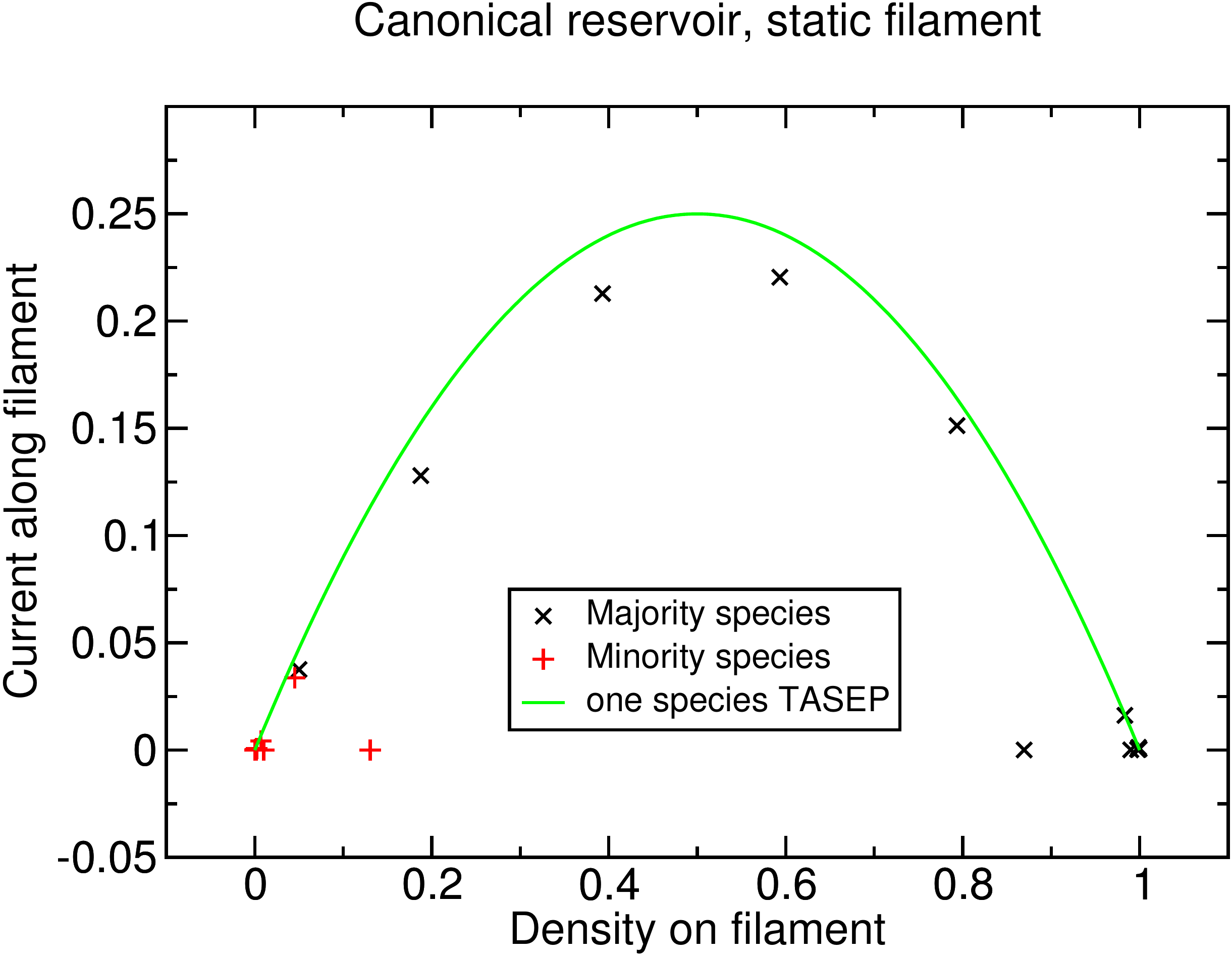}
    \caption{Comparison of the current in a one species TASEP (green line) and
the recorded current along one given filament in the two-species system with a
canonical reservoir and a static lattice,
and with an interaction strength $q=20$ (which ensures that
lane formation has occurred for most densities shown in fig.
\protect{\ref{fig:c_static:magnetization_current}}). The dominating particle
species is shown as black crosses, the other species as red plus signs.
Following the black crosses from left to right along the green parabola, the
total density of particles increases from $\rho = 0.1$ to $\rho=4.0$. The value
for $\rho=8.0$ is far from the TASEP solution as lane formation is not
complete. The system size is $L=200$.}
    \protect\label{fig:c_static:currentq20}
  \end{center}
\end{figure}

The current also reflects these two regimes: For $\rho<1$, the current
increases monotonously with increasing interaction strength. At the
polarization transition, the current jumps toward a value very close
to the current of the one-species TASEP, which is the upper-bound for the
flux on the filament (see \fref{fig:c_static:currentq20}).
For densities $\rho\geq 1$, the behavior is distinctively different as the
curves for the current have much more similarity with those found in the
grand-canonical setup and decrease with further increasing interaction.

\begin{figure}[tbp]
  \begin{center}
    \includegraphics[scale=0.25, clip]{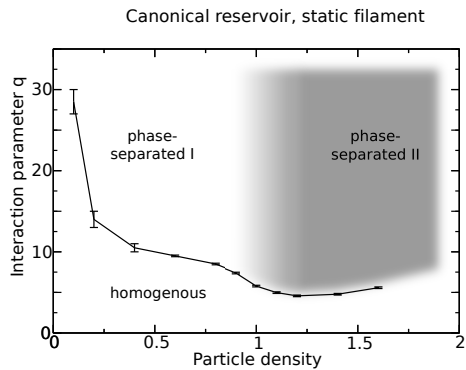}
    \caption{Phase diagram of a system with a canonical reservoir, where $|m|\geq 0.8$
is taken as the criterion for lane separation.
The phase-separated state is
subdivided into two different regions (I and II). In the first region, the
current reaches a constant and large value for strong interactions while the
current decreases in region II, because the filament is overcrowded although
completely polarized.}
    \protect\label{fig:c_static:phase_diagram}
  \end{center}
\end{figure}

On the basis of these observations, we are able to draw an approximative phase
diagram (\fref{fig:c_static:phase_diagram}). The system was considered to be
phase separated at magnetization $|m|\geq 0.8$, where a distinction is made between two different phase-separated regimes depending on the current at infinite interaction strength as explained above. In terms of the transport
capacity, only ``phase-separated I'' can be considered to be an efficient
state, as the current in ``phase-separated II'' does not reach very high values
due to the overcrowding of the filament as discussed above.

In total, we can draw the conclusion that for low densities, the
dominating effect of the interaction is to induce lane formation which
leads to very good values of the current but necessitates strong
particle-particle interactions. For intermediate total densities (greater than one) or a grand-canonical
reservoir, lane formation is not efficient as it occurs at
densities on the filament that are so high that they mostly prevent particles from moving.

\subsubsection{Finite diffusion rate}

\begin{figure}[tbp]
  \begin{center}
    \includegraphics[scale=0.6, clip]{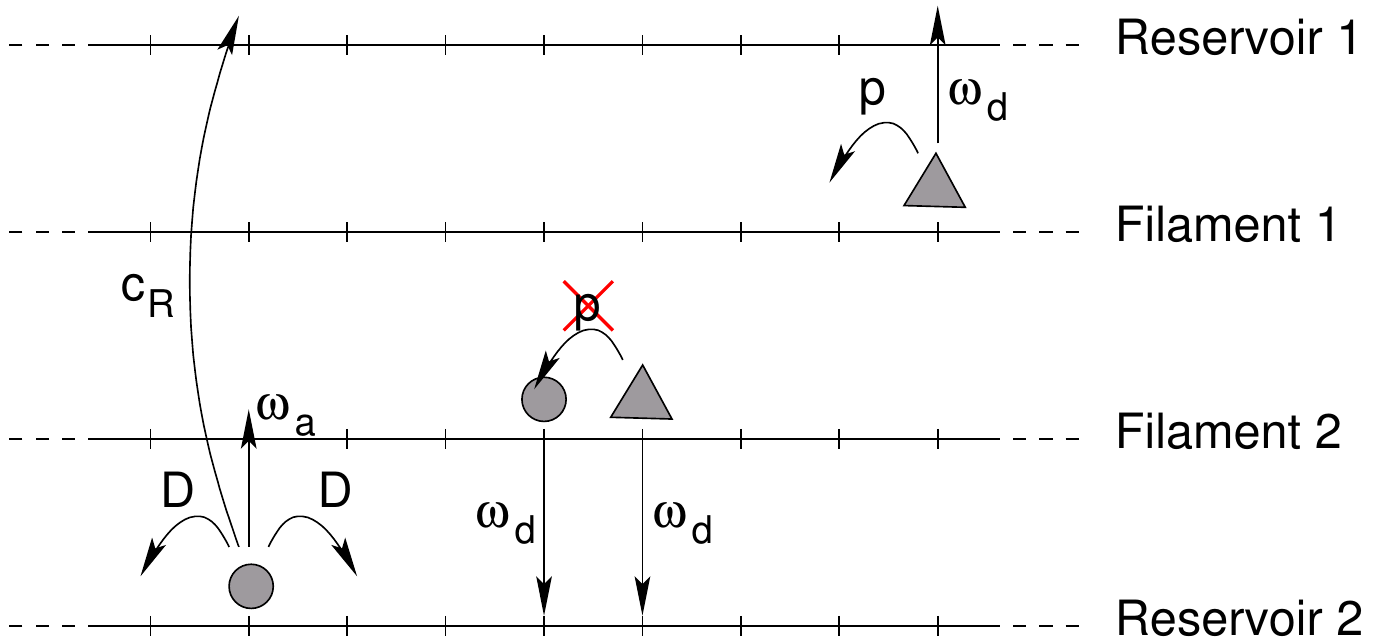}
    \caption{Sketch of the system with two filaments and explicit positions in
the reservoir. In addition to the moves explained for the model with infinite
diffusion (\sref{subsubsec:static_infinite}), we add a coupling between the two
reservoirs by allowing particles to jump from one to the other at rate
$\cR$.}
    \protect\label{fig:2lanemodel}
  \end{center}
\end{figure}

In order to study the effect of confinement, we consider
now a reservoir such as the one defined in section \ref{subsubsec:finite_diff}.
Actually, we consider two identical subsystems
(see \fref{fig:2lanemodel}) for
the same reasons as explained above. Particles can hop from one reservoir to
the other while conserving their longitudinal position at rate $\cR$.
This corresponds to diffusion in transverse direction.
We choose the diffusion in both directions along the reservoir lattices
to be equal to the attachment rate, i.e.,
$D=\cR=\oma=0.1$.
The two filament lattices are not directly coupled in this model, an assumption that facilitates lane formation for this setup.

Without the modified attachment/detachment rates defined in \sref{subsec:interact_def},
this model has been shown to generically form a single big cluster which
contains almost all of the particles in the system~\cite{Ebbinghaus09}. The
existence of this big cluster is a clear indicator for an inefficient state
since it blocks most of the transport. Furthermore, the cluster formation
depends on the number of particles rather than density so that systems form
larger and more stable clusters with increasing system size although the total
density of particles is constant. As a consequence, the current vanishes
in the thermodynamic limit.

\begin{figure}[tbp]
  \begin{center}
    \includegraphics[scale=0.25, clip]{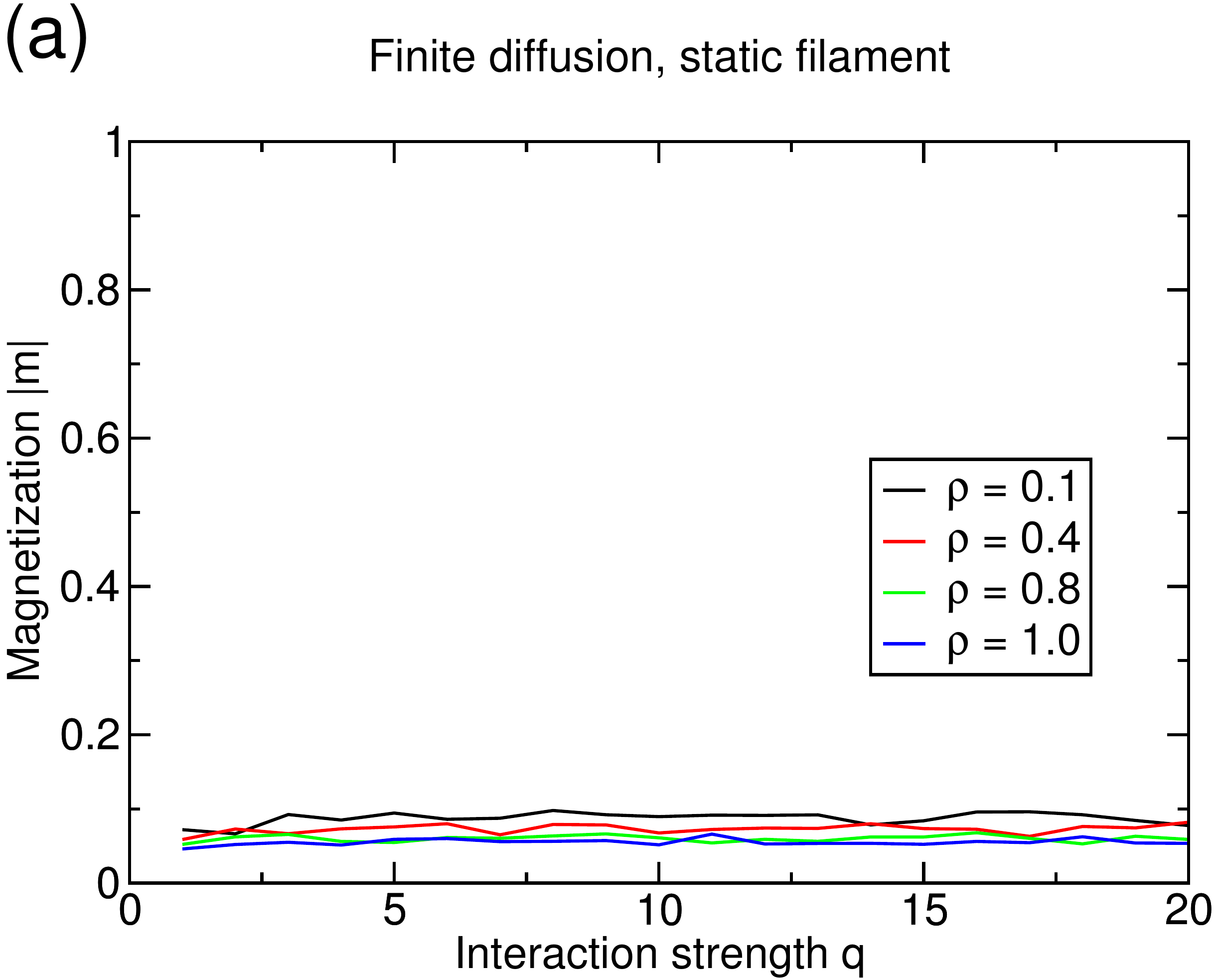}
    \includegraphics[scale=0.25, clip]{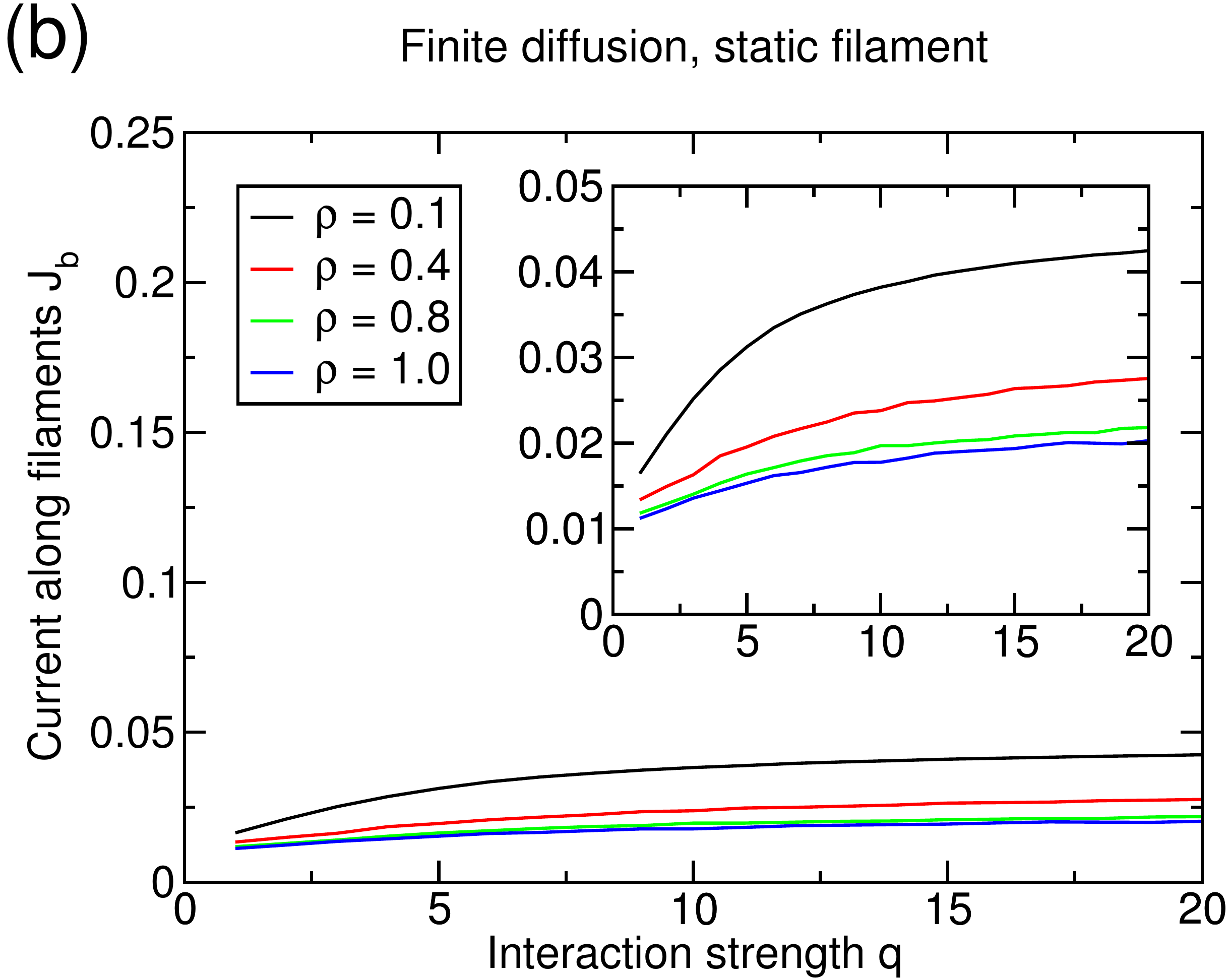}
    \caption{(a) Magnetization of the filament and (b) current of
    positive particles along both filaments in the setup with finite
    diffusivities and a static filament, for a system size $L=1000$ and
    different densities $\rho$.
All quantities are plotted against the interaction strength
$q$ tuning the modified attachment/detachment rates.
    }
    \protect\label{fig:fd_static:magnetization_current}
  \end{center}
\end{figure}

If modified attachment/detachment rates are included, we see that the finite
diffusion in the reservoir inhibits even partial demixing of the particles into
the different subsystems (see \fref{fig:fd_static:magnetization_current}(a)).
Although lane formation is suppressed by the limited diffusion, the
current (\fref{fig:fd_static:magnetization_current}(b)) is enhanced by the
interaction, though a single big cluster is still present in the system. The
increase in the current is caused by the reduced time blocked particles spend in
front of each other and the fact that, in low density regions, particles
of a given species are more likely to have nearest neighbors of the same
species.

\subsection{Dynamic lattice}

In this section, we are interested in the impact that a dynamic
lattice has, first, on the lane formation
mechanism that exists at least in the case of infinite diffusion,
and, second, on the overall transport efficiency.
We shall consider for the filament the simplified dynamic
lattice defined in \sref{subsec:latticedyn_def}.
We again discriminate between reservoirs with infinite and finite
diffusion rates.

\subsubsection{Infinite diffusion rate}\label{subsubsec:KL_dynamic_inifinite}
\begin{figure}[tbp]
  \begin{center}
    \includegraphics[scale=0.25, clip]{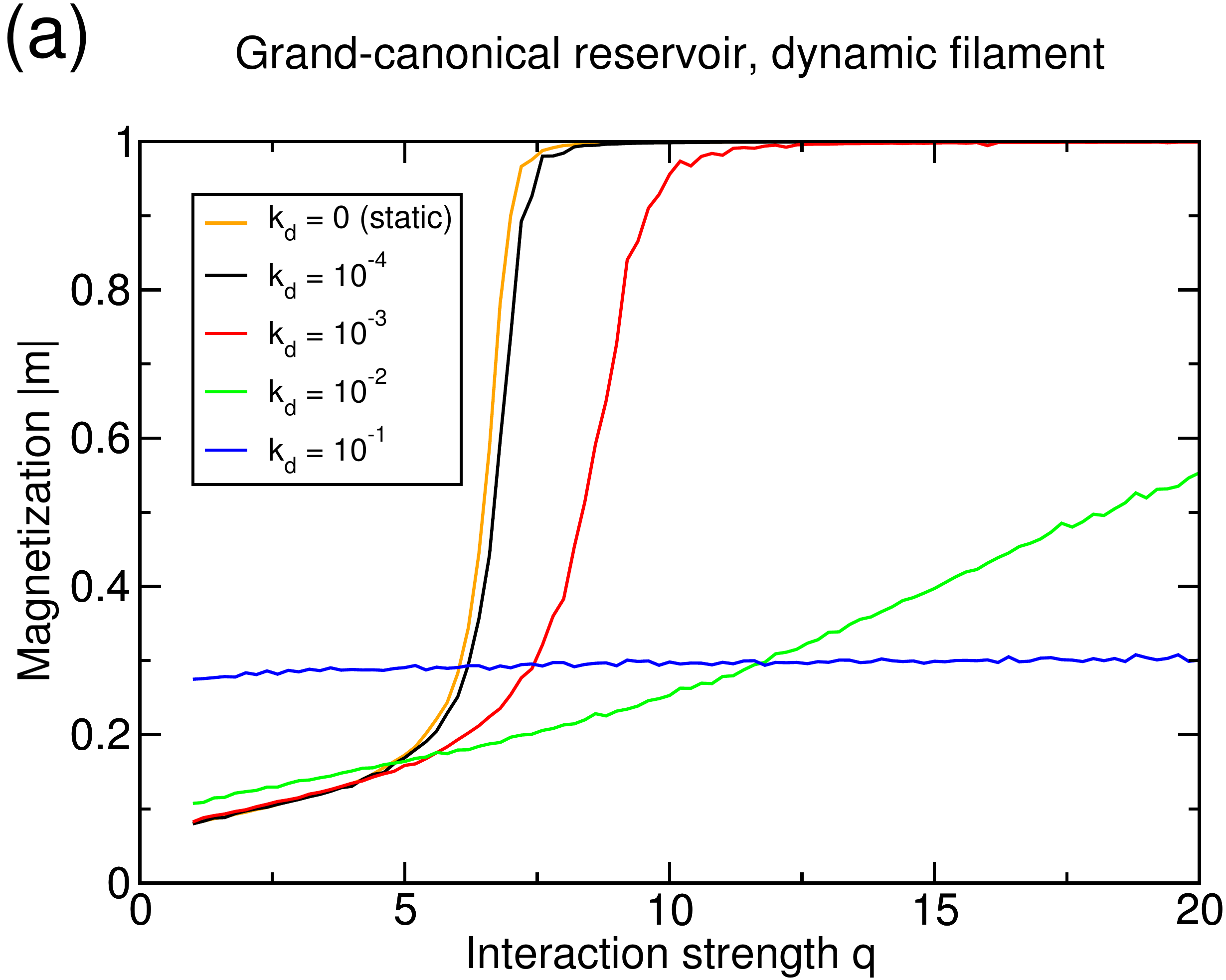}
    \includegraphics[scale=0.25, clip]{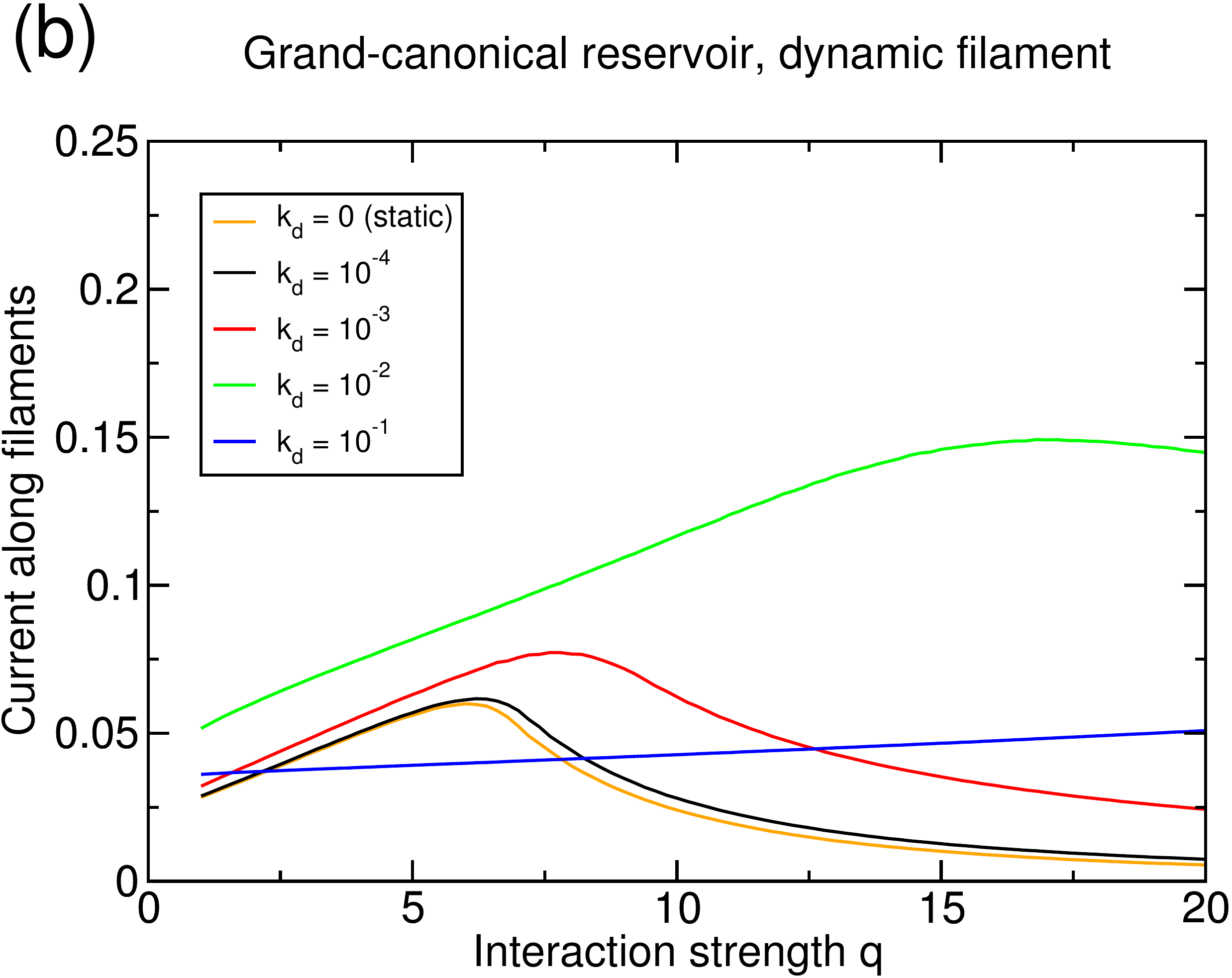}
    \caption{(a) Magnetization of the filament and
    (b) current of positive particles along both
filaments in the grand-canonical setup on a dynamic
lattice of size $L=200$.
The polymerization rate of the dynamic lattice is kept
constant and equal to $k_p=1$.
All quantities are plotted against the interaction strength
$q$ tuning the modified attachment/detachment rates.
The lattice dynamics suppresses lane formation and improves the
current if the lattice dynamics is not too strong.}
    \protect\label{fig:gc_dynamic:magnetization_current}
  \end{center}
\end{figure}

Eliminating lattice sites reduces the effectivity of the particle-particle
interaction since it is next-neighbor and does not reach across holes in the
filament. As a consequence, the transition is shifted to higher values of $q$ with
increasing depolymerization rate $\kd$ as can be seen in
\fref{fig:gc_dynamic:magnetization_current}(a). For a very dynamic lattice,
there seems to be no transition toward lane formation at all. This effect is partly due to the fact that lattice dynamics cuts the filament in shorter segments on which lane formation is less stable, as has been verified by comparison with a static lattice with the same density of holes. However, the effect observed here is stronger if the lattice additionally is dynamic.

However, although lane formation is hindered by the lattice dynamics, it
positively affects the current of the system
(\fref{fig:gc_dynamic:magnetization_current}(b)). By pushing particles to the
reservoir when a lattice site is eliminated, the lattice dynamics keeps the
density on the filament low enough so that particles have enough space in
front of them to move.
Thus, extreme densities that were strongly limiting the current in the
case of a static lattice (see section \ref{subsubsec:static_infinite})
are avoided here.

The results shown in \fref{fig:gc_dynamic:magnetization_current} are obtained for the grand-canonical reservoir but are qualitatively the same for a canonical setup at high densities. For densities $\rho<1$, the current on the static lattice is already close to optimal (see \sref{subsubsec:static_infinite}) and so there is little space for improvement. Actually, the lattice dynamics corrupts these states by shifting the transition to higher values of $q$ and making the phase separation less stable.

\subsubsection{Finite diffusion rate}

In the model with finite diffusion rates in the reservoir, there is no lane
formation on the static lattice and the same remains true if a lattice dynamics
at any strength is added. But as has already been reported
in~\cite{Ebbinghaus10}, a dynamic lattice is capable of dramatically increasing
the current along the lattice and reaching an efficient state by dissolving large clusters. This effect is
obviously also present in this system and the strength of the additional interaction only
has very little impact on the current on the dynamic lattice.

The increase of current due to the modified attachment/detachment rates on the dynamic lattice is of the same order as the one observed on the static lattice (figure~\ref{fig:fd_static:magnetization_current}) and represents only a small fraction of the increase due to lattice dynamics: In a system of density $\rho=1$ and length $L=1000$, the maximum current increase caused by the lattice dynamics is almost ten-fold, whereas the modified attachment/detachment rates do not even double the current in the system.

\subsection{Summary}
\begin{table}[tbp]
\centering
\includegraphics[scale=0.3, clip]{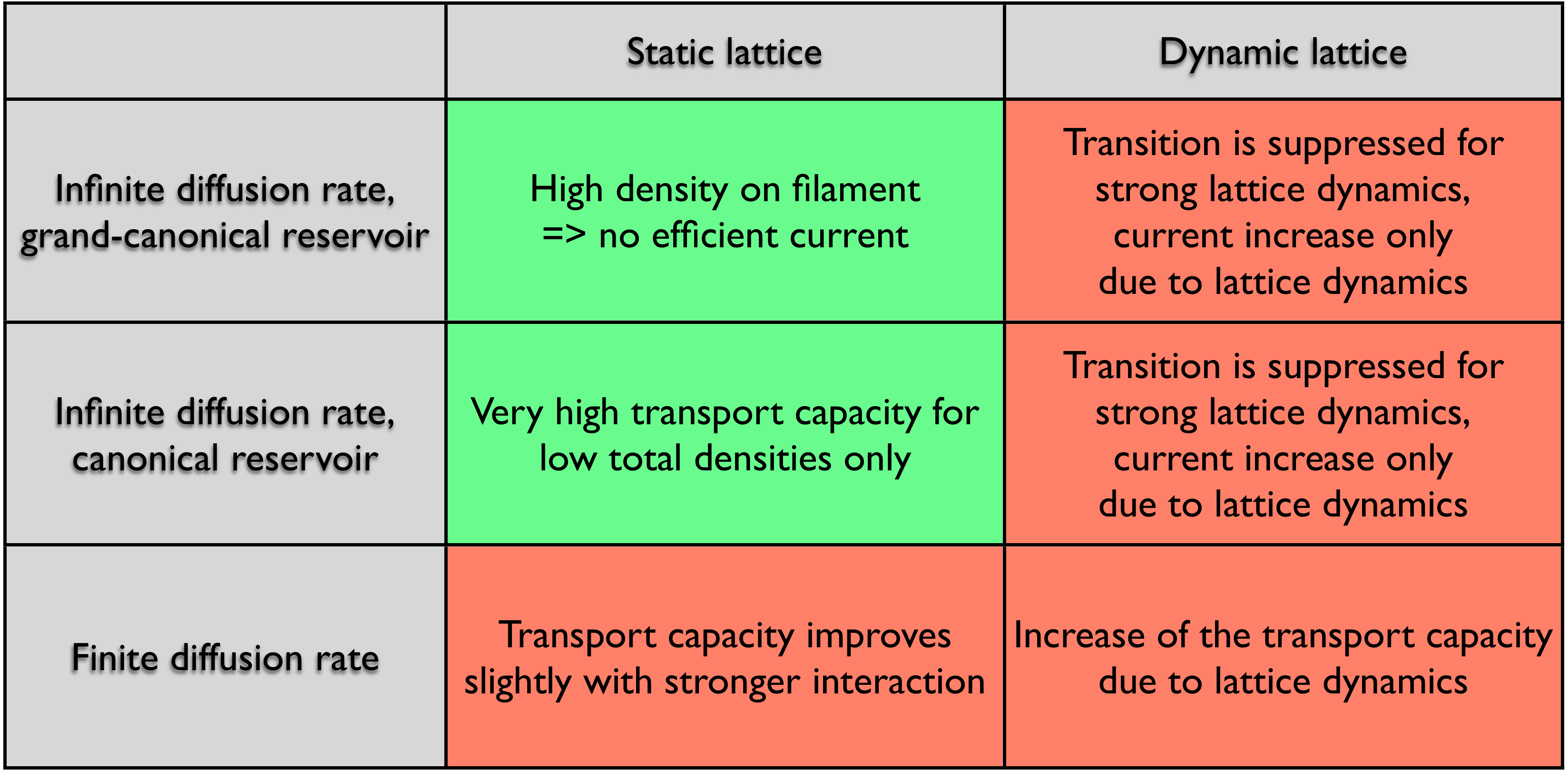}
\caption{\label{tab:KL_results}Overview of results for modified attachment/detachment rates. As a function of the two dimensions \emph{type of reservoir} and \emph{lattice dynamics}, the most important results on the lane formation are given. A green cell indicates that lane formation occurs whereas the inverse is true for a red cell. It should be kept in mind
however that lane formation does not necessarily imply that transport
is efficient, and that efficient transport can also be obtained
without lane formation.
The text in the cell gives additional information on the phenomenology.}
\end{table}

An overview of the results of this section with its multiple scenarios is shown in table~\ref{tab:KL_results} with the color of the cell indicating if lane formation is possible.

In summary, particle-particle interactions which modify the attachment and
detachment rates of particles depending on the occupation of the neighboring
sites can lead to lane formation for reservoirs with an infinite diffusion rate
only. In this case, although the number of encounters of particles of different
species is reduced, the throughput of particles is extremely low in systems
with a {\em large} number of particles. This is due to overcrowding of the lattice
caused by the strong attractive interactions in lane-separated systems.

On the other hand, we have observed that 
efficient transport through lane formation can be obtained if the total
particle density is limited to {\em low} enough values.
However, in order to be in this optimal regime, the interaction strength $q$ should be
large enough. There is no direct experimental measurement of the interaction
strength that would allow to decide whether real systems are in this efficient
regime. Crude estimates for $q$ can be made from the decoration experiments
with kinesin, and would give $q\approx 4$ \cite{Muto05} or $q\approx 2.25$
\cite{Roos08}, thus values that would be too low to trigger lane formation.
However, this is certainly not a definite statement, as the interaction terms could have a different form than the one of figure~\ref{fig:KL_interaction}, and further experimental
measurements (in particular involving two types of motors) should be performed
to refine the prediction.
It should be noted that due to the dynamics of microtubules
in real systems,
the interaction strength $q$ needed to obtain separated lanes
may be shifted to higher values,
as we have found that it is the case here for a much simplified
lattice dynamics.
More generally, the fact that the lane formation
mechanism is dependent on an infinite diffusivity around the transport lattice
is a major drawback for the relevance of this type of interaction for the axon, which is a rather crowded environment.
Thus, though an efficient regime was evidenced by our simulations
for the canonical reservoir,
we are not able to decide yet whether it is relevant for real
systems, and thus it is worth to explore alternative scenarios
(as will be done in the next sections).

In the absence of the aforementioned particle-particle interaction, the lattice dynamics was shown to improve the current along the filament for all three types of reservoirs.
With the interaction presented here, it has a positive effect whenever current is hindered by a (possibly local) high density. The lattice dynamics dominate the behavior of the system in the sense that the lane formation disappears for strong lattice dynamics whereas for strong particle-particle interactions, the effect of the lattice dynamics remains visible.

\section{Side-stepping in front of obstacles}
\label{sec:sidestep}
The interaction presented in the previous section was able to induce
lane formation and efficient transport only in the case
of infinite diffusion in the reservoir.
By contrast, in this section, we propose another type of particle interaction that is capable of inducing lane formation even for finite diffusion rates in the reservoir.
As a consequence, only the reservoir with a finite diffusion rate will be treated here.

\subsection{Model definition}
\begin{figure}[tbp]
  \begin{center}
    \includegraphics[scale=0.8, clip]{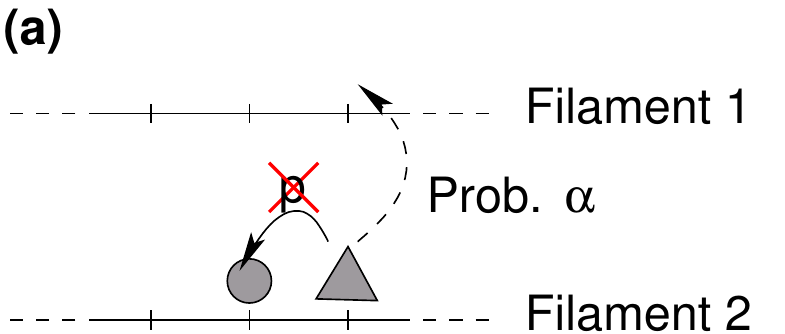}
    \hspace{1.0cm}
    \includegraphics[scale=0.8, clip]{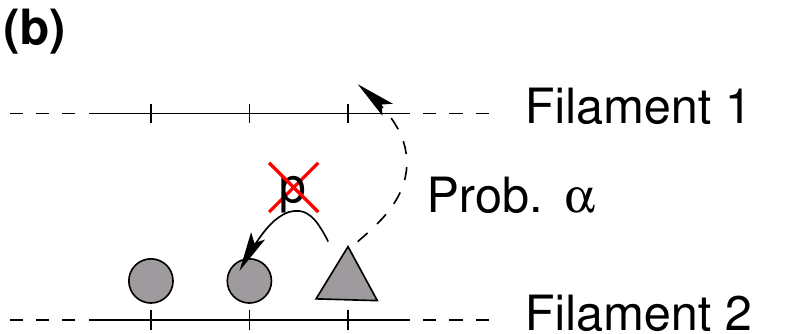}
    \caption{Lane changes (or side stepping) through steric interactions. A particle may change onto the neighboring filament with filament $\alpha$ if it encounters (a) one or (b) two particles of the other species. The reservoirs are not shown.}
    \protect\label{fig:sidestepmodel}
  \end{center}
\end{figure}

Particles obey to the same hopping rules as depicted in \fref{fig:2lanemodel},
i.e., they hop into a preferential direction along the filaments and hop with
equal rates in both directions in the reservoirs.
In order to keep the model simple, we consider again only two filaments and
the two associated reservoirs.
 Exchanges between reservoirs or
reservoir and filament occur with the same rates as before
(see fig.~\ref{fig:2lanemodel}). The particles still
interact via hard-core exclusion on the filament but instead of modifying
attachment and detachment, we consider the following interaction between
particles: If a particle's step along the filament is rejected because
the target site is occupied by a particle of the opposite species, the particle
switches to the other filament with probability $\alpha$ provided the site at
the same longitudinal position is empty on the other filament (see
\fref{fig:sidestepmodel}(a)).
We shall refer to this interaction as the ``single obstacle''
interaction in the following.

So far, there is no experimental data showing a similar behavior of molecular motors. However, the only known experiments on interactions between molecular motors are the decoration experiments for kinesins addressed in section~\ref{sec:interact} which was shown to be rather questionable as a cause for possible lane formation.
Here, the idea is that opposed motors push against each other and thus could exert a force which might induce lane changes. This effect might be even more pronounced if one thinks of the relatively large cargos carried by the molecular motors.

We shall also consider a variant of this interaction
where the 
particle may only switch from one
filament to the other with probability $\alpha$ if it encounters \emph{two}
particles of the opposite species on the two next sites in stepping direction
(see \fref{fig:sidestepmodel}(b)).
This mechanism will be referred to as the ``two obstacles'' interaction.

\subsection{Static lattice}
\subsubsection{``Single obstacle'' interaction}\label{subsubsec:static_single_obst}
On a static lattice, changing lanes in front of a single particle of opposite species does not lead to dramatic effects. In particular, no lane formation is observed although the absolute magnetization of the filament slightly increases. This effect competes with the reservoir exchange rate $\cR$ which tends to mix the system and destroys any phase separation between the two subsystems. At $\cR=D=0.1$, no effect of the additional particle-particle interaction is visible at any probability $\alpha$.

Nevertheless, there is a small effect for decoupled reservoirs at $\cR=0$. The increase of the magnetization is stronger the fewer particles we have in the system, but even for densities as low as $\rho=0.1$, the magnetization does not exceed $|m|<0.3$. It is noteworthy that at the system size $L=1000$ considered here, there is no important clustering at this low density. In a clustered system with higher particle density, absolute increases in magnetization are almost negligible: For $\rho=1$, the magnetization passes from $|m|=0.047$ at $\alpha=0$ to $|m|=0.0784$ at $\alpha=1$.

Alike, the current does not improve much with this side-stepping interaction. The clustered systems remain clustered and exhibit a maximal gain of their current of $25\%$ while starting from an extremely low value being caused by the existence of the large cluster.

\subsubsection{``Two obstacles'' interaction}
If we consider a stronger condition for filament changes as shown
in~\fref{fig:sidestepmodel}(b), lane formation is obtained much more easily
compared to the single obstacle model (\sref{subsubsec:static_single_obst}) and
occurs at any density with decoupled reservoirs, i.e., $\cR=0$. The current of
particles along each filament is consequently very high and close to the
optimal value found in a one-species TASEP. The interaction is asymmetric in the sense that a single particle changes the filament if opposed to two particles of opposite charge while the single particle cannot induce any lane change on the two particles. As a consequence, a particle species which is only slightly dominant on one filament can push more and more particles from the minority species to the other filament. Ultimately, complete lane formation is observed.

\begin{figure}[tbp]
  \begin{center}
    \includegraphics[scale=0.25, clip]{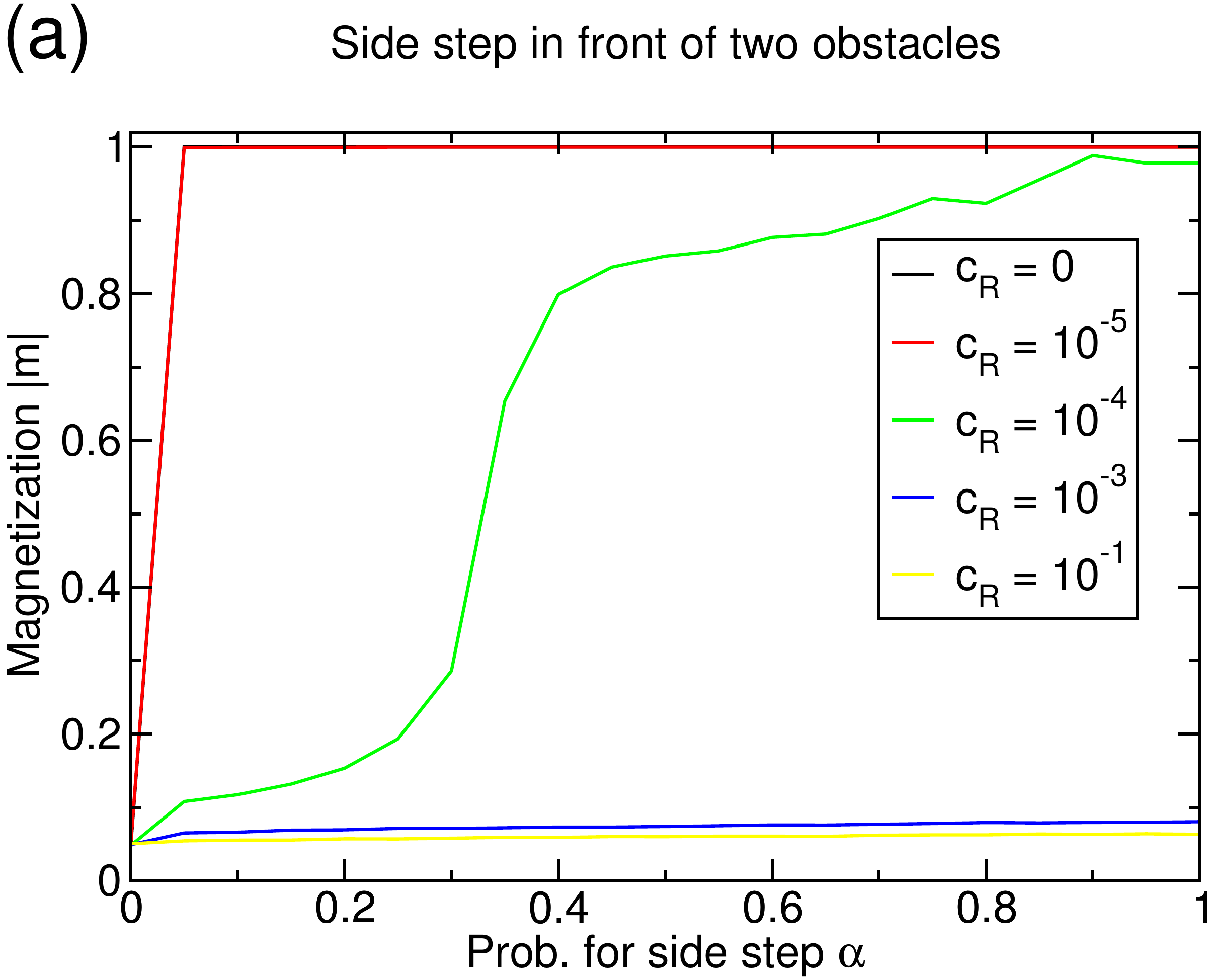}
    \includegraphics[scale=0.25, clip]{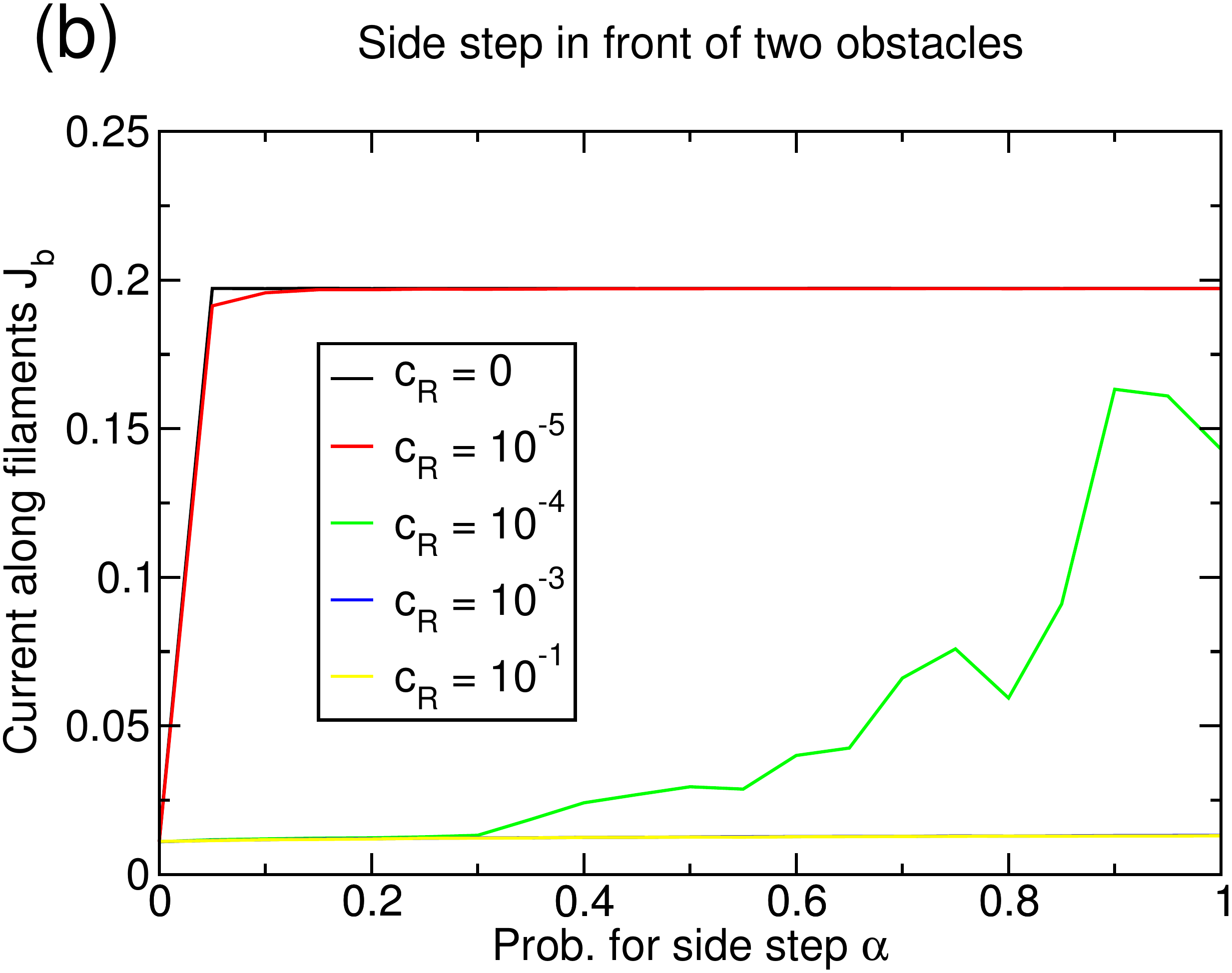}
    \caption{(a) Magnetization of the filament and (b) current along both
    filaments of positive particles on a static lattice with filament changes
    of particles as depicted in \fref{fig:sidestepmodel}(b) (side stepping
    induced by ``two obstacles''). The system size is $L=1000$
    and the total density $\rho=1$. The reservoir
    coupling inhibits lane formation. Systems with lane formation show a
    close-to-optimal current which is one order of magnitude higher than in the
    mixed case.}
    \protect\label{fig:sidestep2_static:magnetization_current}
  \end{center}
\end{figure}

As discussed in the last section, the lane formation competes with the mixing
through the reservoirs. If the reservoir coupling is increased, the
magnetization is lowered at some point and lane formation disappears
(\fref{fig:sidestep2_static:magnetization_current}(a)). The onset of remixing
the separated lanes occurs at lower couplings $\cR$ for higher densities as in
these cases, more particles are continuously transported to reservoir belonging
to the `wrong' filament. For $\cR=D=0.1$, the system is completely mixed at any
density.

The current increases by one order of magnitude if lane formation occurs
(\fref{fig:sidestep2_static:magnetization_current}(b)) and is very close to the
optimal value in the one-species TASEP.

\subsection{Dynamic lattice}
\subsubsection{``Single obstacle'' interaction}\label{subsubsec:dynamic_single_obst}
\begin{figure}[tbp]
  \begin{center}
    \includegraphics[scale=0.25, clip]{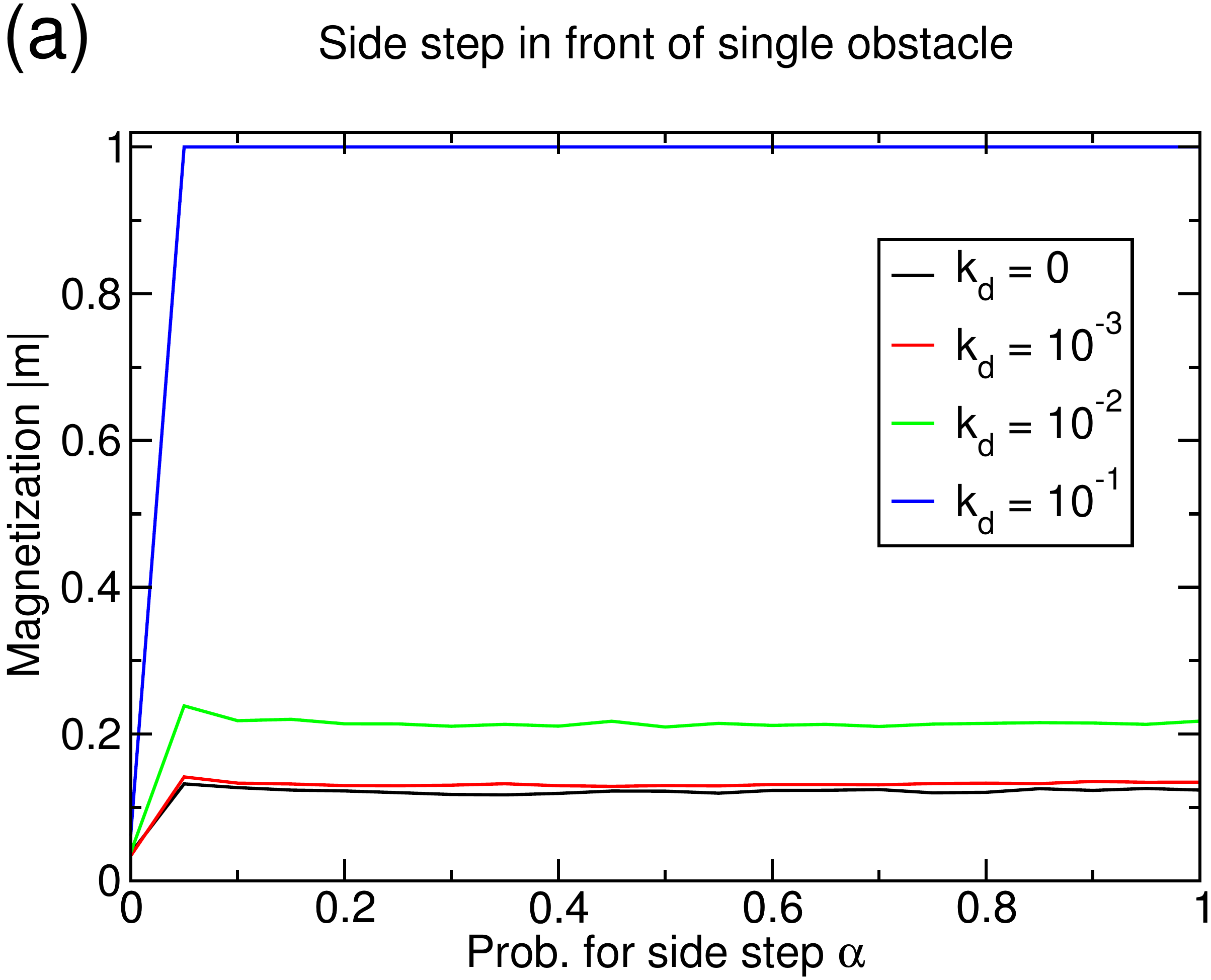}
    \includegraphics[scale=0.25, clip]{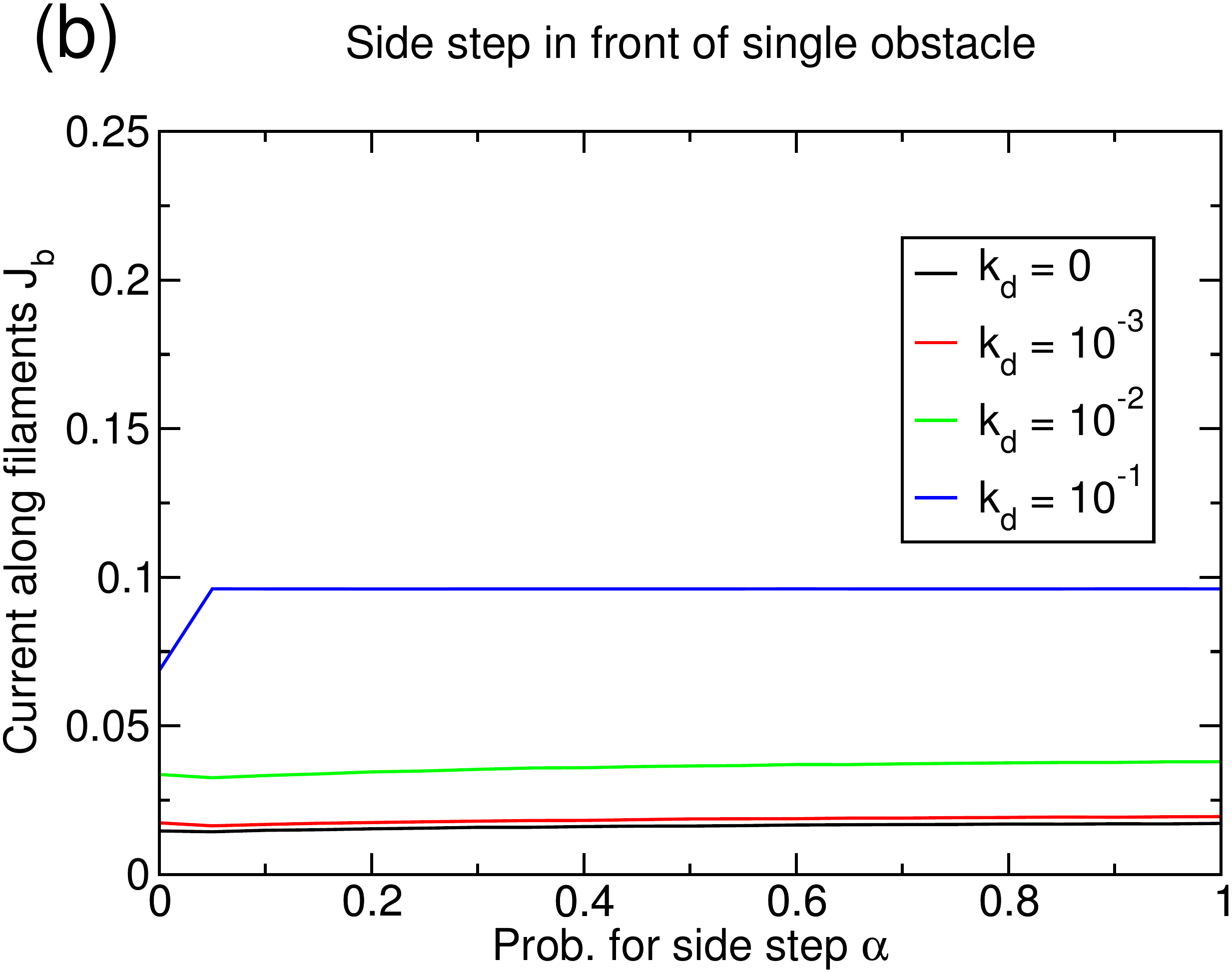}\vspace{5mm}
    \includegraphics[scale=0.25, clip]{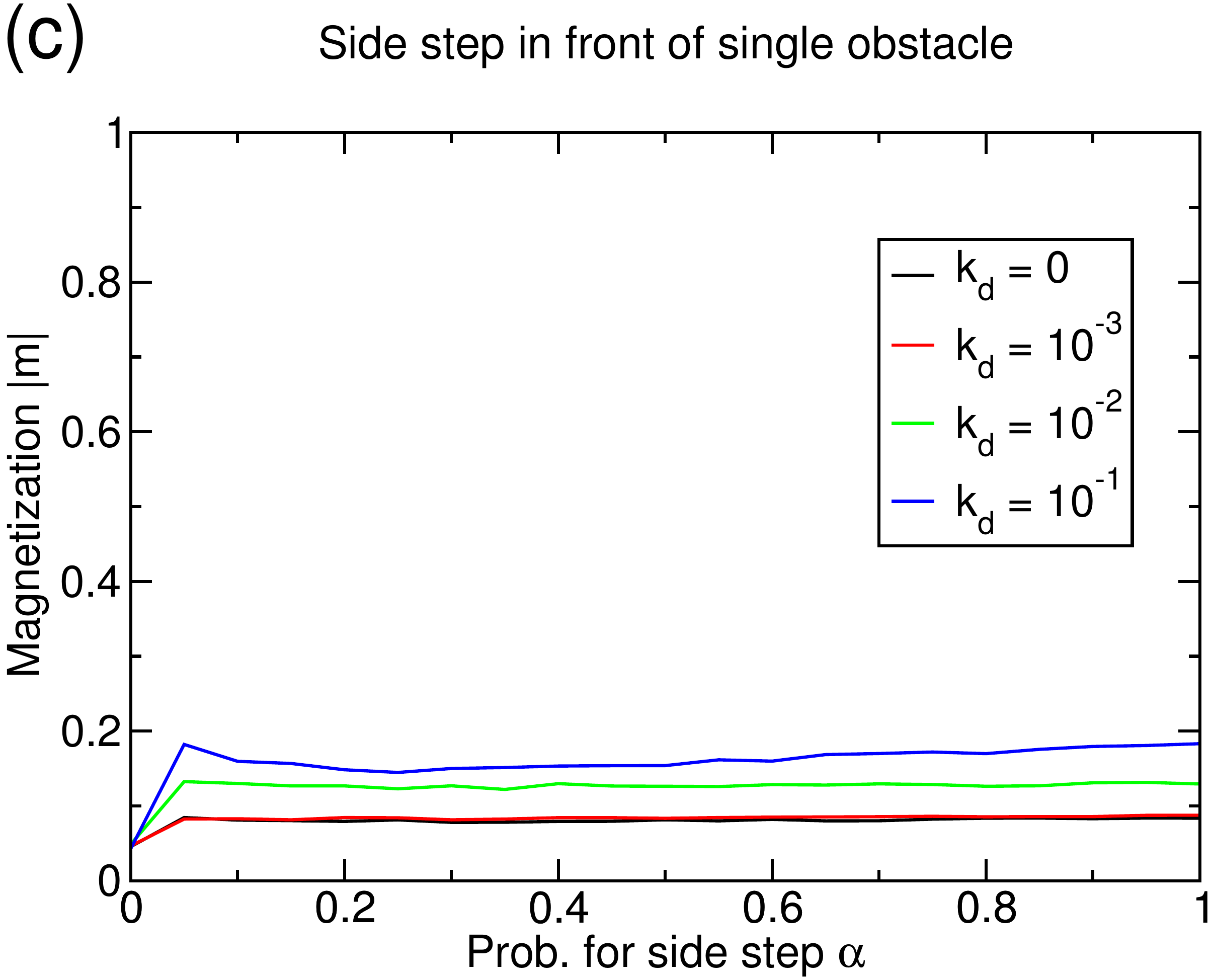}
    \includegraphics[scale=0.25, clip]{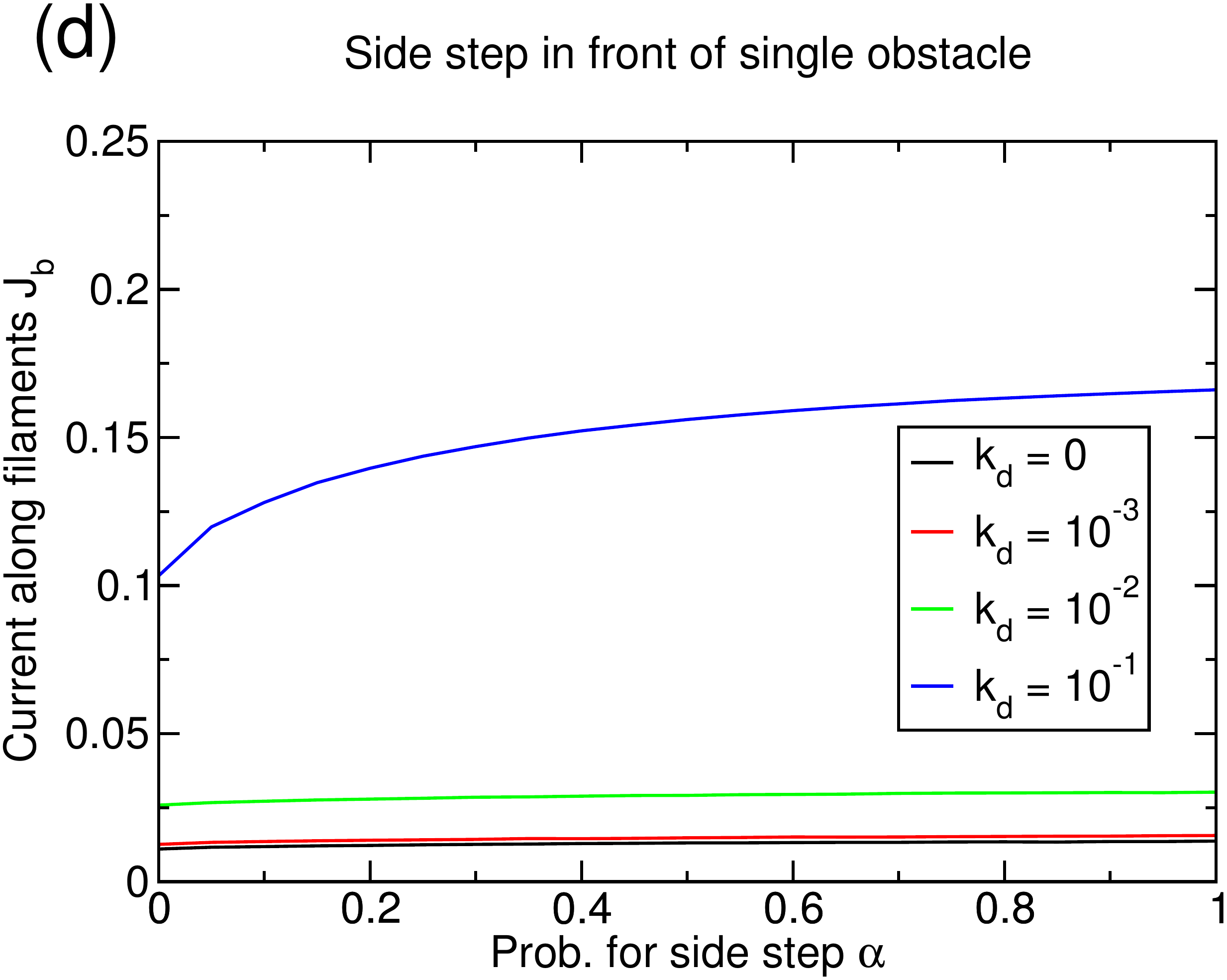}
    \caption{(a,c) Magnetization of the filament and (b,d) current of one
particle species along both filaments in a system of length $L=1000$ with
decoupled reservoirs $\cR=0$, for a density (a,b) $\rho=0.4$ or (c,d)
$\rho=1$ and ``single obstacle'' interaction. For $\rho=0.4$, the enhancement of lane formation
by the lattice dynamics is obvious (a), while it is hardly existent for
$\rho=1$ (c). Lattice dynamics have a stronger impact on the current than the
lane formation as currents in (b) and (d) are comparable although only one of
the systems exhibits lane formation.}
    \protect\label{fig:sidestep1:magnetization_current}
  \end{center}
\end{figure}

If the possibility to perform filament changes is combined with lattice
dynamics, the system's tendency toward lane formation is increased (see
\fref{fig:sidestep1:magnetization_current}(a)). A system of density $\rho=0.4$
reaches complete lane formation at $\kd=0.1$.
However, at this value of the
depolymerization rate $\kd$, the basic model without filament changes does not
exhibit a large cluster anymore, but is in a homogenous state with very small
structures, and thus has already improved transport properties
only due to lattice dynamics.

The enhanced lane formation can be understood in the following way: In order to form lanes, filament changes of particles are necessary which can only occur if particles of different species meet. Consequently, more lane changes are induced if there are more interfaces between positive and negative particles. In a clustered system, almost all the particles are accumulated in the cluster so that there are only very few interfaces. The lattice dynamics, though, inhibits the formation of large clusters and instead distributes the particles more homogeneously over the whole system leading to many very small clusters. This effectively increases the number of interfaces between positive and negative particles and thus favors lane changes and lane formation. The effect which has been mentioned in
\sref{subsubsec:KL_dynamic_inifinite} that holes in the lattice decrease the
effectivity of a particle-particle interaction because the interaction is only
next-neighbor seems to be overcompensated by the increase in interfaces due to
the dissolution of the big cluster.

The strength of the lattice dynamics that is needed to form lanes
 depends on the particle density, e.g., a system of
density $\rho=1$ does not exhibit complete lane formation at any lattice
dynamics (see \fref{fig:sidestep1:magnetization_current}(c)) while a system of
density $\rho=0.1$ forms lanes at much lower lattice dynamics (not shown).

Again, the current is more strongly (and positively) affected by the lattice dynamics than by the
additional particle-particle interaction as can be seen in
\fref{fig:sidestep1:magnetization_current}(b) and (d): The higher the lattice
dynamics, the higher the overall position of the current curves. In particular,
for the completely lane separated system at $\rho=0.4$ for $\alpha>0$, the
current increases by $40\%$ due to the lane formation, but more than quadruples
because of the lattice dynamics
(\fref{fig:sidestep1:magnetization_current}(b)). This impression is
strengthened by the observation that similar gains due to the dynamic lattice
are obtained even if no lane formation occurs
(\fref{fig:sidestep1:magnetization_current}(d)). The values of the current for
$\rho=1$ are remarkable since they are rather close to the maximum current of
$1/4$ that would be found in a one-species TASEP, although encounters between
particles of different species are still frequent in our model even at
important lattice dynamics.

\begin{figure}[tbp]
  \begin{center}
    \includegraphics[scale=0.25, clip]{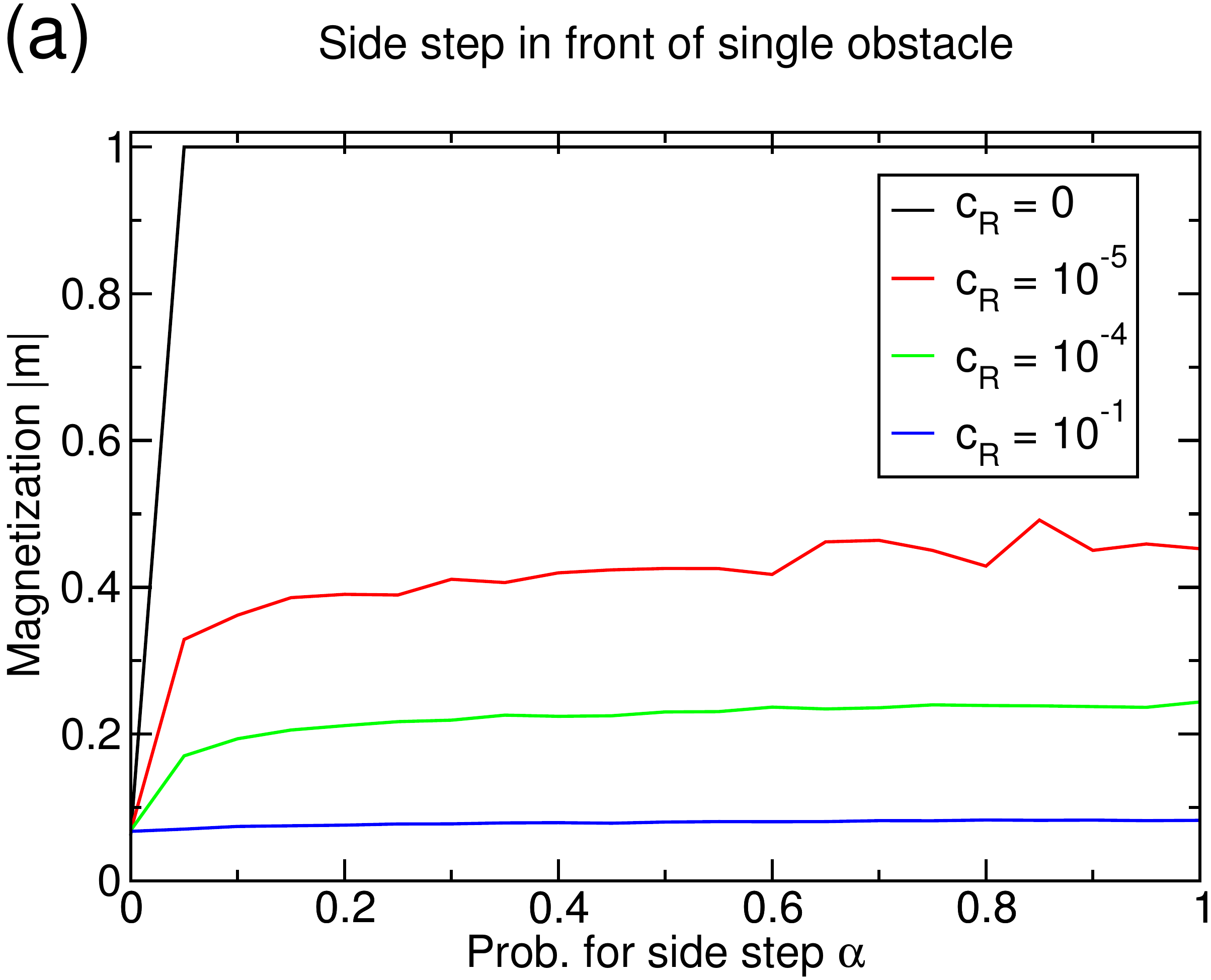}
    \includegraphics[scale=0.25, clip]{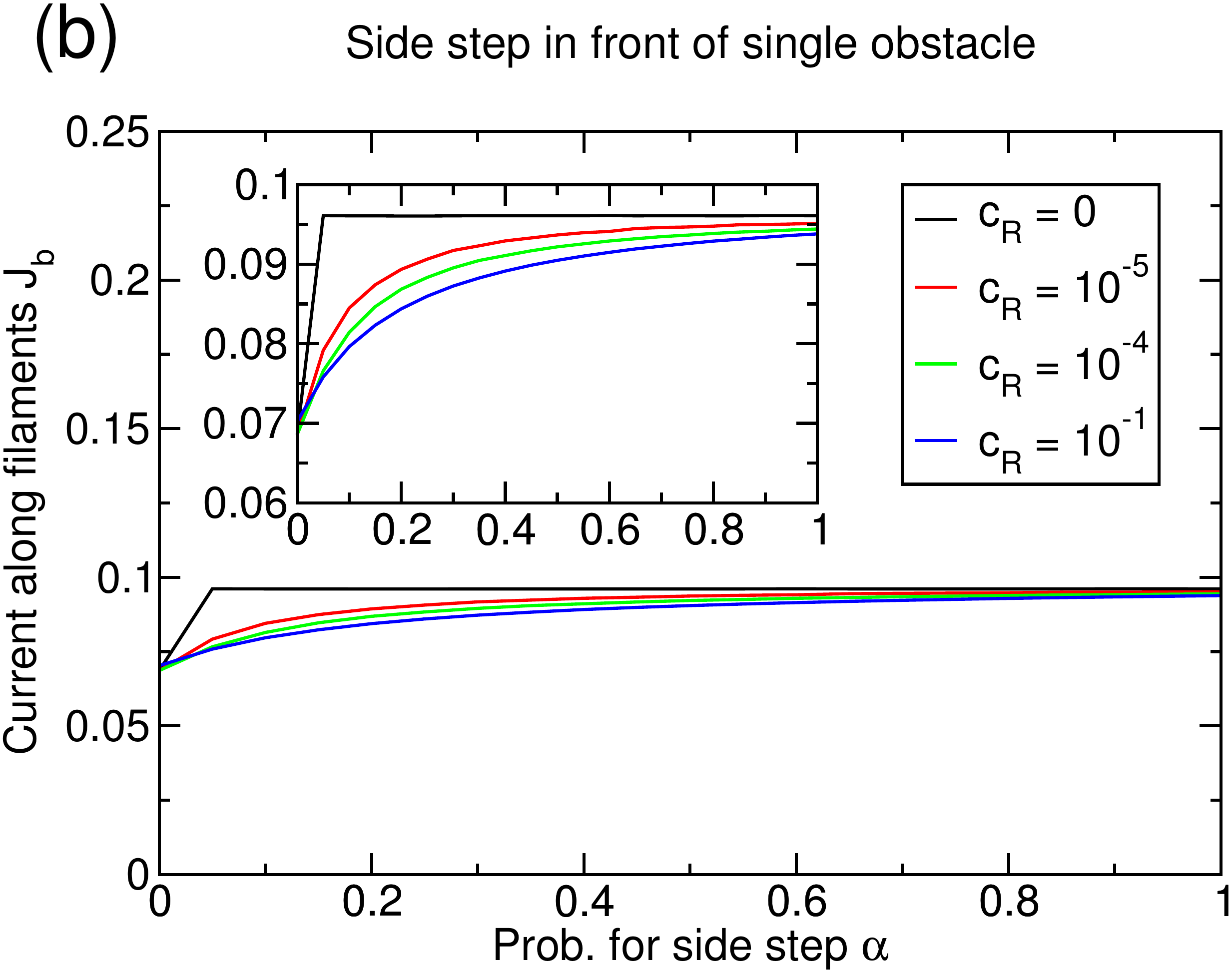}
    \caption{(a) Magnetization of the filament and (b) current of one particle species along both filaments in a system of length $L=1000$ at density $\rho=0.4$. Motors interact through the ``single obstacle'' interaction, and a dynamic lattice is considered with $k_d=0.1$.
    The different curves correspond to different reservoir couplings which acts against lane formation induced by the ``single obstacle'' interaction.}
    \protect\label{fig:sidestep1:magnetization_current_cR}
  \end{center}
\end{figure}

These results have so far been obtained at decoupled reservoirs $\cR=0$. The influence of the reservoir coupling $\cR$ is shown in
\fref{fig:sidestep1:magnetization_current_cR} for a dynamic lattice. Already an extremely small
reservoir coupling destroys the lane formation
(\fref{fig:sidestep1:magnetization_current_cR}(a)). If diffusion in
longitudinal and transverse direction are equal, i.e., $\cR=D=0.1$, the lane
formation effect disappears almost completely. However, the current is only
marginally affected (\fref{fig:sidestep1:magnetization_current_cR}(b)) and
reaches almost identical values for all reservoir couplings at $\alpha=1$. This
once more confirms that the lattice dynamics is dominant in determining the
transport capacity of the system.

\subsubsection{``Two obstacles'' interaction}
\begin{figure}[tbp]
  \begin{center}
    \includegraphics[scale=0.25, clip]{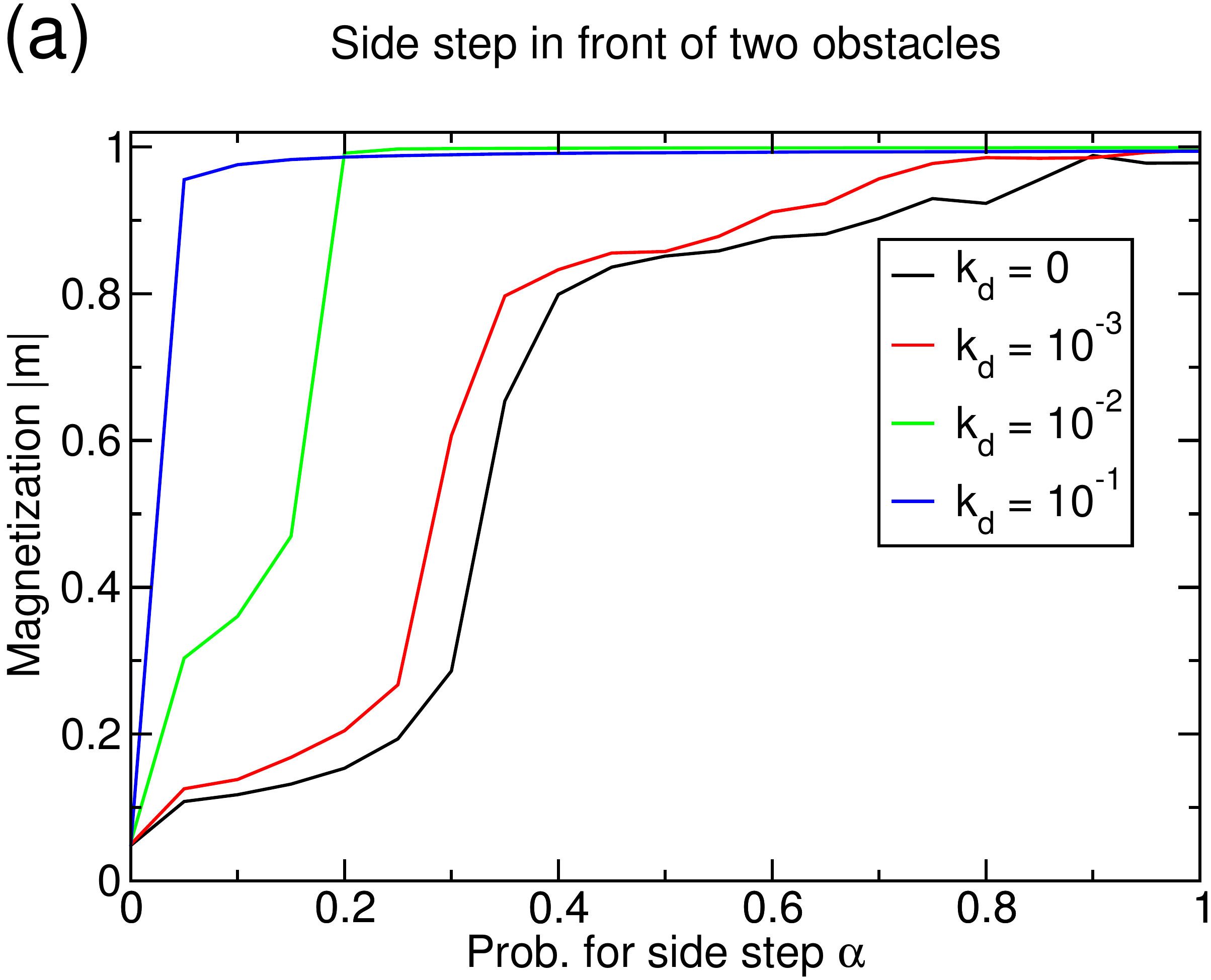}
    \includegraphics[scale=0.25, clip]{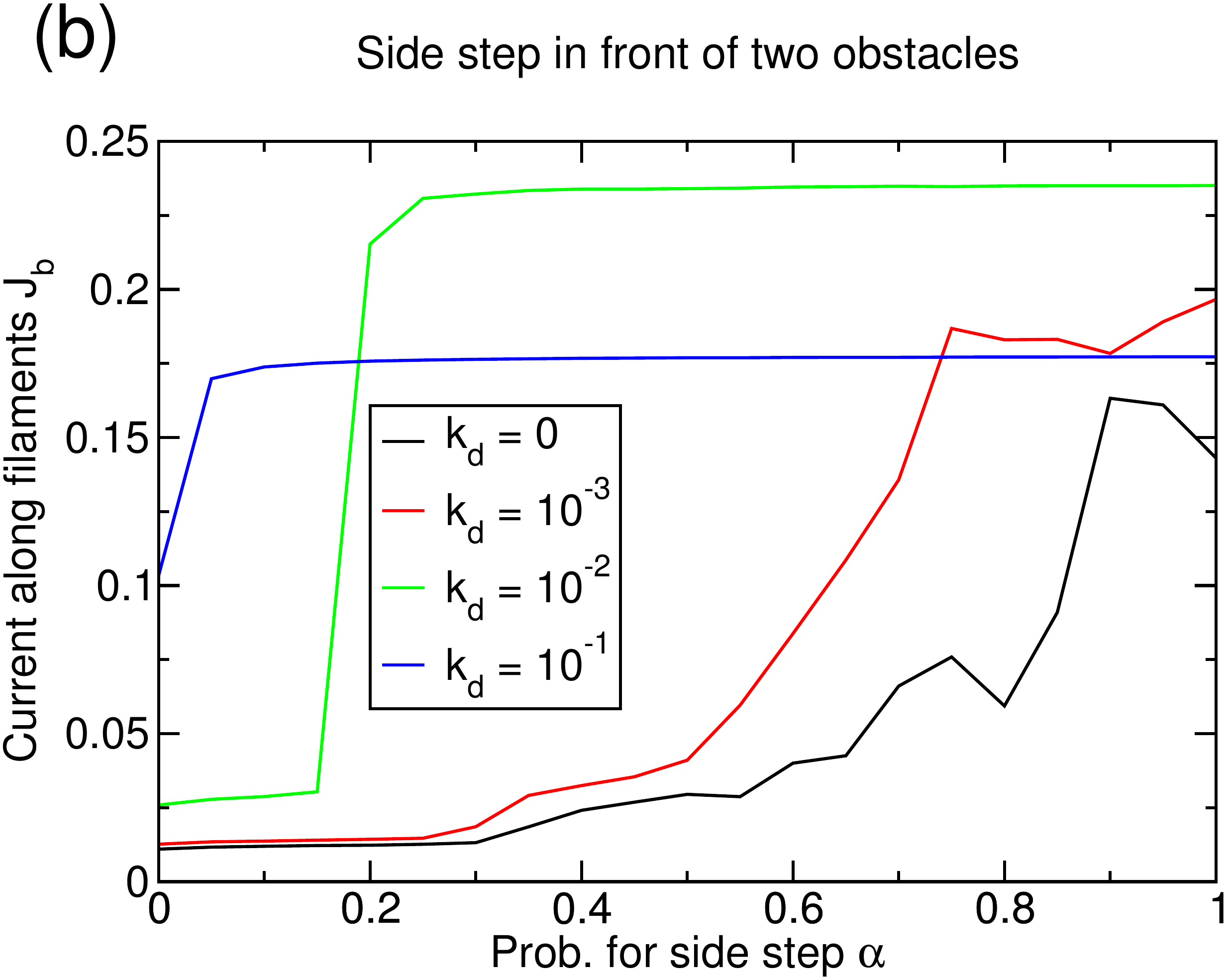}
    \caption{(a) Magnetization of the filament and (b) current along both
    filaments of positive particles on a dynamic lattice of size $L=1000$ at density $\rho=1.0$,
    with ``two obstacle'' interactions
 as depicted in \fref{fig:sidestepmodel}(b), and reservoir
    coupling $\cR=10^{-4}$. The lattice dynamics enhance lane formation in only
    partially de-mixed systems and have an important impact on the current.}
    \protect\label{fig:sidestep2_dynamic:magnetization_current}
  \end{center}
\end{figure}

As for the the ``single obstacle'' interaction, the lattice dynamics also improve the lane
formation mechanism by increasing the number of interfaces between oppositely
moving particles for the ``two obstacles'' interaction (\fref{fig:sidestep2_dynamic:magnetization_current}(a)). The
influence of the lattice dynamics and the competition with the reservoir
coupling $\cR$ on the magnetization and therefore on the lane formation remains
qualitatively unchanged compared to \sref{subsubsec:dynamic_single_obst}.

There is nevertheless an interesting difference for the value of the current in
the systems with lane formation: For the most dynamic lattices $\kd=0.1$, the
current in the separated system can sometimes be lower than in de-mixed or even
mixed systems at lower lattice dynamics (blue and green lines in
\fref{fig:sidestep2_dynamic:magnetization_current}(b)). In these cases, the lattice dynamics
actually inhibit any further improvement of the current by eliminating too many
sites of the filament along which no transport can happen. This effect is the
more pronounced, the lower the density of particles is.

\subsection{Summary}

\begin{table}[tbp]
\centering
\includegraphics[scale=0.3, clip]{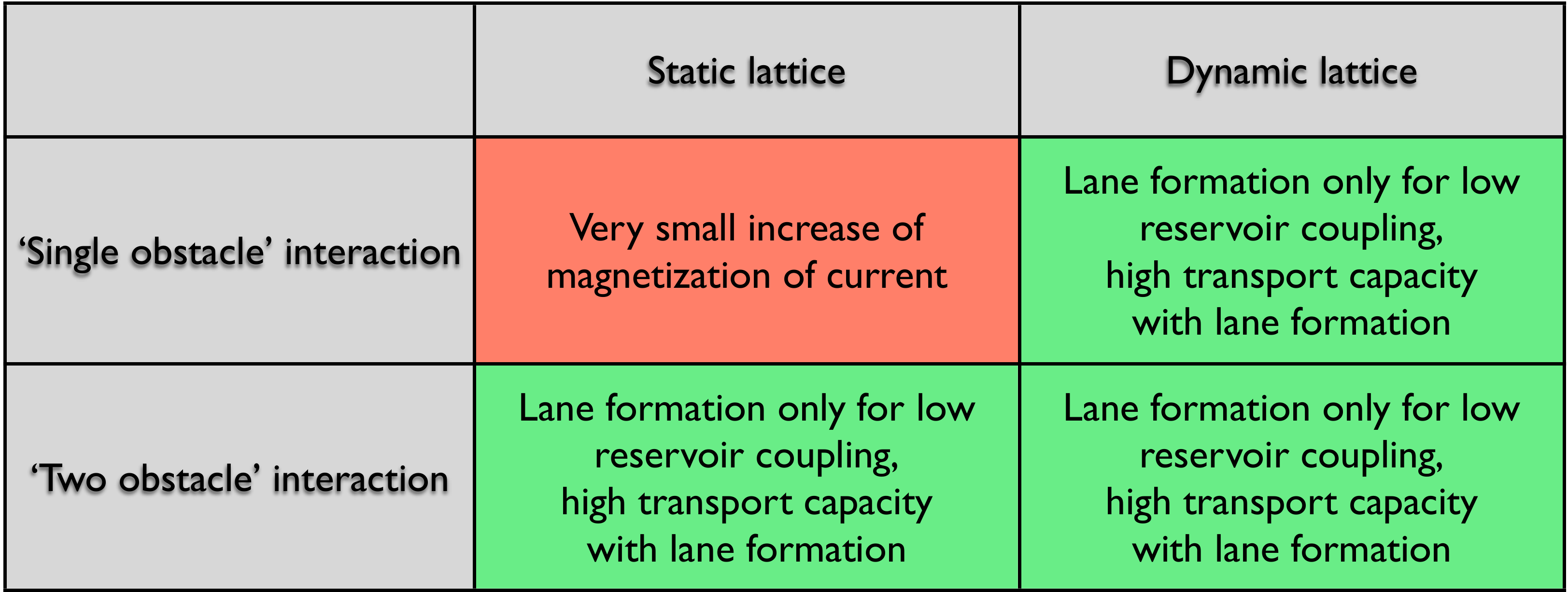}
\caption[Overview of results for lane formation through direct filament changes.]{\label{tab:stericIA_results}Overview of results for lane formation through direct filament changes. As a function of the two dimensions \emph{type of interaction} and \emph{lattice dynamics}, the most important results on the lane formation are given. A green cell indicates that lane formation occurs whereas the inverse is true for a red cell. The text in the cell gives additional information on the phenomenology.
Note that for a dynamic lattice, high transport capacities
are maintained even for higher reservoir couplings, though
lane formation is hindered.}
\end{table}

The most important results of this section are summarized in table~\ref{tab:stericIA_results}.

The proposed interaction between particles which eventually induces filament
changes of the particles has been shown to lead to lane formation under special
conditions. In general, three factors determine the possibility of lane
formation: 1. A higher particle density $\rho$ decreases the tendency to form
separate lanes for both species (in contrast to what was observed for the
modified attachment/detachment rates of section \ref{sec:interact}). 2. The reservoir coupling $\cR$ suppresses
lane formation by mixing partially phase-separated subsystems (already true
for the modified attachment/detachment rates of section \ref{sec:interact}). This mixing is
very efficient as the rates needed to inhibit lane formation are several orders
of magnitude smaller than all the other rates in the system. 3. Lattice
dynamics promotes lane formation by increasing the number of contacts between
positive and negative particles through a redistribution of the particles on several filament segments. We therefore have two effects leading to
higher currents, one of which (lattice dynamics) improves the other (lane
formation).

Considering the current, lane-separated states can obviously be considered as
efficient transport states. They have a high throughput of particles and
because of the absence of large structures, no finite-size effects are
present, so that we have a density-dependent state. This property is
necessary to have finite current in very long systems. However, we have shown
that the effect of lattice dynamics on the current has a much larger amplitude
than the one due to the side stepping mechanism.

\section{Discussion}
In the present article, we presented two different types of particle-particle
interactions in a quasi one-dimensional lattice gas with oppositely moving
particle species. Both types of interactions lead to lane formation under
certain conditions. We explored the robustness of these lane-separated
states and their transport capacity. An efficient transport
state was supposed to be characterized by a system size-independent (=
density-dependent) state, with a finite current that compares to the current in
a one-species TASEP serving as a benchmark.

The first type of interaction which was considered by Klumpp and Lipowsky~\cite{Klumpp04} affects the affinity to the filament in the vicinity of other particles. The lane-separated states that are found (only for infinite reservoir diffusion) exhibit a density-dependent current but many of these states also have extremely high densities on the filament so that the actual current of bound particles is rather low compared to an optimal regime. This is in particular the case for open systems (grand-canonical reservoir), or for closed systems with high enough densities (canonical reservoir).
The conclusion is that lane formation does not necessarily imply efficient transport although this can be the case.
Lower densities with a canonical reservoir do lead to very efficient states which compare well to the one-species TASEP.

Lane formation with this type of interaction (leading to modified
attachment/detachment rates) was only possible for an infinite diffusion rate
in the reservoir. With regard to an application of these concepts to real
transport processes as intracellular transport, this is a very strong
assumption. The interior of a cell is very crowded and considering the typical
sizes of vesicles and organelles, diffusion should be very limited.
In the case of finite diffusion, no lane formation is observed and
the interaction plays a minor role in improving the current, compared
to the much more positive impact of lattice dynamics.

In the second type of interaction that has been considered in this paper, particles hop to the neighboring filament when they meet an oppositely charged motor. Although this model is in principle able to exhibit lane formation at finite longitudinal diffusion, the lane-separated state is extremely sensitive to diffusion between reservoirs, i.e., diffusion in the transverse direction. This is in sharp contrast to the first interaction model which actually needed infinite diffusion rates to exhibit lane formation. The difference comes from the nature of the particle-particle interaction.  Any de-mixing in the first scenario has to be
carried out via particle exchanges with the reservoir whereas in the second scenario, particles directly switch from one
filament to the other and do not need the reservoir to form lanes.
From a biological point of view, the strong limitation of transversal diffusion required for the scenario
based on side-stepping could be realized if the axon is
compartmentalized such that diffusing particles cannot reach easily other
filaments than the ones in their direct neighborhood.

Furthermore, based on the experimental observation that
microtubule-based transport in axons takes place on dynamic
microtubules, we have considered that the underlying lattice itself can
undergo some dynamics, through the much simplified random
polymerization/depolymerization of filament units.
As, from a general point of view, the consequences of a lattice
dynamics on transport properties are not known,
we chose to concentrate on a simplified lattice dynamics,
to study generic effects,
rather than going directly into the full complexity of the
biological system.

We had already shown in \cite{Ebbinghaus10} how such a lattice dynamics
can drive the system from a size-dependent state with inefficient transport, to a density-dependent state with efficient transport.
In the present paper, we have studied again what would be the influence of a dynamic lattice on the transport efficiency, but now in the presence of interactions.
In almost all cases (i.e., for almost all types of reservoirs and particle interactions),
we found that lattice dynamics improves the efficiency
of transport. 
Actually, we found only one case in which the
lattice dynamics does not improve the current in the system: if the  system
forms lanes at low enough density $\rho$, with infinite diffusion rate in the reservoir,
then lattice dynamics worsens
a state which is already close to optimal transport.
However, an infinite diffusion rate in the reservoir is not expected to be relevant
for transport in real cells, and besides, there is no experimental
evidence yet that the motor-motor interaction strength is sufficient
to reach this state.
 In all the other cases,
the benefits from lattice dynamics as described in~\cite{Ebbinghaus10} were
still valid also when combined with particle-particle interactions.

More precisely, the lattice dynamics modifies the lane formation
mechanism for both types of interaction scenarios. In the first scenario, while lane formation is
shifted to stronger interactions, lattice dynamics still improve the transport
in the system by lowering the density on the filament in the inefficiently
crowded states. The second lane formation mechanism through side-stepping is enhanced by the
lattice dynamics as it relies on direct contacts between particles of opposite
species which are increased by the dissolution of clusters.
Then lattice dynamics on the one hand lowers the interaction
strength required to form lanes, and on the other hand
improves the current in lane separated states.

For both types of interactions, the same effects are already present on a static lattice with a sufficient density of holes, due to an effective limitation of the system size.
But our results indicate that the effects are enhanced when the lattice
undergoes some dynamics (at equal hole density).

In total, lane formation appears difficult to achieve in bidirectional
systems as considered here, because the fact that the run length
along the filament is finite (due to processivity or lattice dynamics)
actively works against any sorting of particles onto different filaments.
Lane formation can be achieved only for special
combinations of parameters or characteristics of the system.
It is an open question to know whether biological parameters
for axonal transport are or are not in this regime.
Several types of molecular motors which participate in
intracellular transport have been studied experimentally, but
results are obtained mostly for \emph{in vitro} experiments, while it
is not obvious that dynamical parameters are the same in an in
vivo environment.  Besides, some motors (as dyneins) are
more difficult to study, and we are lacking data. There are
almost no experimental results about interactions between
different types of motors.
It cannot be excluded that other interactions between motors than those considered here lead to lane formation mechanisms which are less sensitive to system parameters. There is an urgent need for more data on particle-particle interactions which could possibly be acquired by statistical exploitation of \emph{in vitro} experiments involving many motors.

As mentioned above, the lattice dynamics were intentionally
 kept simple in this paper. We found that such lattice dynamics
 produce transport enhancement through cluster
 dissolution, and in \cite{Ebbinghaus10}, we found that it was robust
 for several
 types of simple lattice dynamics. Still, in further studies, we plan
 to apply our approach to more biologically relevant dynamics,
 but there is still a need of experimental observations
about the precise dynamics of the microtubule network.

Anyhow, for the interactions considered in this paper, we found that for
finite diffusion the lattice dynamics always
has a positive effect on transport, whether lane formation
is present or not.
Thus in the context of axonal transport, lattice dynamics
and motor-motor interactions should not
be considered as competing mechanisms, but 
rather as two complementary processes that can contribute
in a cooperative way to the efficiency of bidirectional transport.
Only further experimental informations can allow to quantify
the contribution of the various ingredients.
However, in view of our results,
we suggest that the microtubule dynamics could play an important
role in the regulation of bidirectional transport, whether
in combination with motor-motor interactions or not.

\ack
ME would like to thank the DFG Research Training Group 1276 for financial support.

\section*{Bibliography}
\bibliographystyle{unsrt}


\end{document}